\documentclass[aps,pre,notitlepage,tightenlines,12pt,nofootinbib]{revtex4-1}

\pdfoutput=1

\usepackage{ifthen}
\def\journalversion{false}
\newcommand{\ifjournal}[2]{\ifthenelse{\equal{\journalversion}{true}}{#1}{#2}}
\newcommand{\coloronline}{\ifjournal{(Color online)\ }{}}

\usepackage{amsmath}
\usepackage{amssymb,amsfonts}
\usepackage{amsthm}
\usepackage{graphicx}
\usepackage{subfigure}
\usepackage{bm}
\usepackage{color}
\usepackage{hyperref}

\graphicspath{{figs/}}

\DeclareMathOperator{\sinc}{sinc}
\DeclareMathOperator{\arctanh}{arctanh}

\newtheorem{prop}{Proposition}
\theoremstyle{definition}

\newcommand{\ie}{\textit{i.e.}}
\newcommand{\etal}{\textit{et~al.}}
\newcommand{\mathnotation}[2]{\newcommand{#1}{\ensuremath{#2}}}

\newcommand{\Order}[1]{\mathrm{O}(#1)}      
\newcommand{\order}[1]{\mathrm{o}(#1)}      
\newcommand{\avg}[1]{\langle #1 \rangle}    
\newcommand{\indf}[1]{\l[#1\r]}             

\newcommand{\citeauthoreos}[2]{\citeauthor{#1}\ }

\renewcommand{\l}{\left}
\renewcommand{\r}{\right}
\mathnotation{\ldef}{\mathrel{\raisebox{.069ex}{:}\!\!=}}
\mathnotation{\ee}{\mathrm{e}}              
\mathnotation{\imi}{\mathrm{i}}             

\mathnotation{\dint}{\,{\mathrm{d}}}        
\mathnotation{\Uc}{U}                       
\mathnotation{\Uv}{\bm{\Uc}}                
\mathnotation{\uv}{\bm{u}}                  
\mathnotation{\pal}{\lambda}                
\mathnotation{\lsc}{\ell}                   
\mathnotation{\nd}{n}                       
\mathnotation{\Vol}{V}                      
\mathnotation{\Volint}{\widetilde{\Vol}_\pal}
\mathnotation{\nenc}{m}                     
\mathnotation{\Nswim}{N}                    
\mathnotation{\sdim}{d}                     
\mathnotation{\Diff}{D}                     
\mathnotation{\X}{X}                        
\mathnotation{\Rc}{R}                       
\mathnotation{\Rv}{\bm{\Rc}}                
\mathnotation{\xc}{x}                       
\mathnotation{\yc}{y}                       
\mathnotation{\zc}{z}                       
\mathnotation{\rc}{r}                       
\mathnotation{\rv}{\bm{\rc}}                
\mathnotation{\xv}{\bm{x}}                  
\mathnotation{\etav}{\bm{\eta}}             
\mathnotation{\zetav}{\bm{\zeta}}           
\mathnotation{\Deltav}{\bm{\Delta}}         
\mathnotation{\Deltasc}{D}                  
\renewcommand{\tt}{t}                       
\mathnotation{\stime}{s}                    
\mathnotation{\prob}{p}                     
\mathnotation{\probone}{p}                  
\mathnotation{\Reff}{R_{\text{eff}}}        
\mathnotation{\areasw}{\sigma}              
\mathnotation{\mfp}{\ell_{\text{mfp}}}      
\mathnotation{\Mpal}{\nu_\pal}              
\mathnotation{\Mpalm}{\tilde{\nu}_\pal}     
\mathnotation{\vpal}{v_\pal}                
\mathnotation{\alsct}{\sqrt{\pal}}          
\mathnotation{\K}{\gamma}                   
\mathnotation{\cK}{\Gamma}                  
\mathnotation{\cKm}{\widetilde{\Gamma}}     
\newcommand{\kc}{k}                         
\mathnotation{\kmax}{\kc_{\text{max}}}      
\mathnotation{\areaus}{\Omega}              
\mathnotation{\Clog}{C}                     
\mathnotation{\bb}{b}                       

\mathnotation{\volsw}{v_{\text{sw}}}        
\mathnotation{\I}{I}                        
\mathnotation{\xa}{X}                       
\mathnotation{\cc}{c}                       
\mathnotation{\Lpal}{\Lambda}               
\renewcommand{\sc}{s}                       

\mathnotation{\yprop}{y}
\mathnotation{\Mprop}{M}
\mathnotation{\eprop}{\eps}
\mathnotation{\eps}{\varepsilon}

\mathnotation{\microm}{\mu\mathrm{m}}
\mathnotation{\second}{\mathrm{s}}

\newcommand{\pdf}{pdf}

\begin{document}

\title{Distribution of particle displacements\\ due to swimming microorganisms}

\author{Jean-Luc Thiffeault}
\email{jeanluc@math.wisc.edu}
\affiliation{Department of Mathematics, University of Wisconsin --
  Madison, 480 Lincoln Dr., Madison, WI 53706, USA}

\begin{abstract}
  The experiments of Leptos~\etal. [\textit{Phys. Rev. Lett.} \textbf{103},
  198103 (2009)] show that the displacements of small particles affected by
  swimming microorganisms achieve a non-Gaussian distribution, which
  nevertheless scales diffusively --- the `diffusive scaling.'  We use a
  simple model where the particles undergo repeated `kicks' due to the
  swimmers to explain the shape of the distribution as a function of the
  volume fraction of swimmers.  The net displacement is determined by the
  inverse Fourier transform of a single-swimmer characteristic function.  The
  only adjustable parameter is the strength of the stresslet term in our
  spherical squirmer model.  We give a criterion for convergence to a Gaussian
  distribution in terms of moments of the drift function, and show that the
  experimentally-observed diffusive scaling is a transient related to the slow
  crossover of the fourth moment from a ballistic to a linear regime with path
  length.  We also present a simple model, with logarithmic drift function,
  that can be solved analytically.
\end{abstract}

\keywords{microswimmers; swimming microorganisms; effective diffusivity;
  particle transport}


\maketitle

\section{Introduction}
\label{sec:intro}

The study of microswimming has exploded in recent years with the advent of
precise, well-controlled experiments.  (See for instance the reviews
of~\citet{Pedley1992} and~\citet{Lauga2009}.) This has uncovered a plethora of
fascinating behavior, for example the complex interaction of microswimmers
with boundaries~\cite{Rothschild1963, Winet1984, Cosson2003, Lauga2006,
  Berke2008, Drescher2009}, or the collective suspension instability (swirls
and jets) at high concentrations of `pushers,' organisms whose propulsion
mechanism is at the rear~\cite{Dombrowski2004, HernandezOrtiz2005,
  Underhill2008, Underhill2011, Saintillan2007, Sokolov2009, Saintillan2012}.

Another fruitful research direction is biogenic mixing, or biomixing for
short.  Does the motion of swimmers influence the effective diffusivity of
passive scalars advected by the fluid, such as the nutrients the organisms
depend on?  This has been proposed as a mixing mechanism in the
ocean~\cite{Huntley2004, Dewar2006, Kunze2006, Katija2009, Dabiri2010,
  Leshansky2010, Thiffeault2010b, Lorke2010, Katija2012}, though its
effectiveness is still very much open to debate~\cite{Visser2007, Gregg2009,
  Kunze2011, Noss2014}.  Biomixing has also been studied in suspensions of
small organisms~\cite{Ishikawa2010, Kurtuldu2011, Mino2011, Zaid2011}.

The main ingredient in formulating a theory for the enhanced diffusion due to
swimming organisms is the \emph{drift} caused by the
swimmer~\cite{Maxwell1869, Darwin1953, Lighthill1956}.  \citet{Katija2009} and
\citet{Thiffeault2010b} proposed that the enhanced diffusivity is due to the
repeated displacements induced by a swimmer on a particle of fluid.
\citet{Thiffeault2010b} and~\citet{Lin2011} formulated a probabilistic model
where, given the drift caused by one swimmer, an effective diffusivity could
be computed.  This model has been tested in physical and numerical
experiments~\cite{Jepson2013, Morozov2014, Kasyap2014} and modified to include
curved trajectories~\cite{Pushkin2013b} and confined
environments~\cite{Pushkin2014}.  Mi\~{n}o
\emph{et~al.}~\cite{Mino2011,Mino2013} observe that effective diffusivity is
inversely related to swimming efficiency, and find increased diffusivity near
solid surfaces, both theoretically and experimentally.  The drift caused by
individual microswimmers has also been studied in its own
right~\cite{Dunkel2010, Pushkin2013}.  \citet{Pushkin2013b} also found an
analytical expression for stresslet displacements, valid in the far field.

The studies mentioned above have typically been concerned with the effective
diffusivity induced by the swimmers, but one can also ask more detailed
questions about the distribution of displacements of fluid particles.
\citet{Wu2000} studied the displacement of spheres larger than the swimming
organisms.  More recently, \citet{Leptos2009} studied the microscopic algae
\textit{Chlamydomonas reinhardtii}.  They used spheres that are much smaller
than the organisms, so their distributions can be taken to be close to the
displacements of idealized fluid particles.  The probability density function
(\pdf) of tracer displacements was found to be strongly non-Gaussian, though
the distributions scaled `diffusively': they collapsed onto each other if
rescaled by their standard deviation.

Several papers have dealt with these non-Gaussian distributions.
\citet{Zaid2011} examine the velocity fluctuations due to swimmers modeled as
regularized point stresslets, and obtain strongly non-Gaussian tails.  The
non-Gaussianity in their case is due to the divergence of the stresslet near
the singularity, which indicates large displacements.  While the broad outline
of this mechanism is surely correct, examining this singular limit is
questionable: it is never valid to evaluate terms such as the stresslet in the
singular limit, since the swimmer's body necessarily regularizes the velocity.
In addition, no direct comparison to experiments is offered beyond a comment
that the data `resemble the measurements of \citet{Leptos2009}.'
\citet{Pushkin2014} extended this work to confined environments, and we will
contrast their results to ours.  As we will show here, the non-Gaussianity
arises from the rarity of interaction events --- the system is very far from
the Gaussian limit.  Note also that \citet{Eckhardt2012} have fitted the
distributions of \citet{Leptos2009} very well to a continuous-time random walk
model, but this does not suggest a mechanism and requires fitting different
parameters at each concentration.

What causes the non-Gaussian form of the displacement distribution?  As was
pointed out by \citet{Pushkin2014}, the experiments are run for a very short
time.  Let us quantify what is meant by `short.'  \citet{Leptos2009} define a
`sphere of influence' of radius~$\Reff$ around a particle: swimmers outside
that sphere do not significantly displace the particle.  If swimmers with
number density~$\nd$ moves a distance~$\pal$ in random directions, the
expected number of `interactions' with a target particle is roughly
\begin{equation*}
  \nd\pal\,\pi\Reff^2
  \sim 0.4.
\end{equation*}
Here we took~$\pal \sim 30\,\microm$ and $\nd \sim 4\times
10^{-5}\,\microm^{-3}$, which are the largest values used in the experiments,
and~$\Reff \sim 10\,\microm$ as estimated in \citet{Leptos2009}.  Hence, a
typical fluid particle feels \emph{very few} near-encounters with any swimmer.
In order for the central limit theorem to apply, the net displacement must be
the sum of many independent displacements, and this is clearly not the case
here for the larger values of the displacement.  We thus expect a Gaussian
core (due to the many small displacements a particle feels) but non-Gaussian
tails (due to the rarity of large displacements), which is exactly what was
observed in the experiments.

Here, we present a calculation that quantitatively predicts essentially all
the details of the distributions obtained by~\citet{Leptos2009}.  The
underlying model is not new, being based on the particle-displacement picture
of \citet{Thiffeault2010b} and~\citet{Lin2011}.  However, the analysis is new:
we show how to combine multiple displacements to obtain the probability
density function due to multiple swimmers, and take the appropriate
infinite-volume limit.  As we go, we discuss the mathematical assumptions that
are required.  Upon comparing with experiments, we find the agreement to be
excellent, in spite of the differences between our model swimmer and the
experiments.  Only a single parameter needs to be fitted: the dimensionless
stresslet strength, $\beta$.

The paper is organized as follows.  In Section~\ref{sec:pdfdisp} we derive the
probability density of displacements based on the drift function of a single
swimmer, in the infinite-volume limit.  We use numerical simulations of a
model swimmer (of the squirmer
type~\cite{Lighthill1952,Blake1971,Ishikawa2006, Ishikawa2007b, Drescher2009})
in Section~\ref{sec:numerics} to obtain a distribution of displacements which
we match to the experiments of \citeauthoreos{Leptos2009}.  In
Section~\ref{sec:interact} we give a different interpretation of the main
formula of Section~\ref{sec:pdfdisp} in terms of `interactions' between
swimmers and a fluid particle.  This alternative form can be used to obtain
some analytic results, in particular when the drift function is logarithmic.
We examine in Section~\ref{sec:largedev} the long-time (or long swimming path)
asymptotics of the model, and find what features of the drift function affect
the convergence to Gaussian.  In Section~\ref{sec:diffscal} we address the
`diffusive scaling' observed in the experiments, and show that it is a
transient phenomenon.  Finally, we discuss our results as well as future
directions in Section~\ref{sec:discussion}.

\section{Distribution of displacements}
\label{sec:pdfdisp}

The setting of our problem is a large volume~$\Vol$ that contains a number of
swimmers~$\Nswim$, also typically large.  The swimmers move independently of
each other in randomly directions.  In the dilute limit that we consider, the
velocity field of one swimmer is not significantly affected by the others.  A
random fluid particle (not too near the edges of the domain), will be
displaced by the cumulative action of the swimmers.  If we follow the
displacements of a large number of well-separated fluid particles, which we
treat as independent, we can obtain the full \pdf\ of displacements. Our goal
is to derive the exact \pdf\ of displacements from a simple probabilistic
model.  Our starting point is the model described by~\citet{Thiffeault2010b}
and improved by~\citet{Lin2011}, which captures the important features
observed in experiments.

For simplicity, we assume the swimmers move along straight paths at a fixed
speed~$\Uc$.  The velocity field induced at point~$\xv$ by a swimmer
is~$\uv(\xv - \Uv\tt)$, with the time dependence reflecting the motion of the
swimmer.  The main ingredient in the model is the finite-path drift
function~$\Deltav_\pal(\etav)$ for a fluid particle, initially at~$\xv=\etav$,
affected by a single swimmer:
\begin{equation}
  \Deltav_\pal(\etav) = \int_0^{\pal/\Uc}
  \uv(\xv(\stime) - \Uv\stime)\dint\stime,\qquad
  \dot\xv = \uv(\xv - \Uv\tt),\quad
  \xv(0) = \etav\,.
  \label{eq:Deltav}
\end{equation}
Here~$\Uc\tt = \pal$ is the swimming distance.  To
obtain~$\Deltav_\pal(\etav)$ we must solve the differential
equation~$\dot\xv=\uv$ for each initial condition~$\etav$.  Assuming
homogeneity and isotropy, we obtain the probability density of
displacements~\cite{Pushkin2014},
\begin{equation}
  \probone_{\Rv_\pal^1}(\rv) = \frac{1}{\areaus\,\rc^{\sdim-1}}\int_\Vol
  \delta(\rc - \Delta_\pal(\etav))
  \,\frac{\!\dint\Vol_{\etav}}{\Vol}
  \label{eq:rhotx}
\end{equation}
where~$\areaus=\areaus(\sdim)$ is the area of the unit sphere in~$\sdim$
dimensions: $\areaus(2)=2\pi$, $\areaus(3)=4\pi$.  Here~$\Rv_\pal^1$ is a
random variable that gives the displacement of the particle from its initial
position after being affected by a single swimmer with path length~$\pal$.  We
denote by $\probone_{\Rv_\pal^1}(\rv)$ the \pdf\
of~$\Rv_\pal^1$.  Because of the isotropy assumption, only the
magnitude~$\Delta_\pal(\etav) = \lVert\Deltav_\pal(\etav)\rVert$
enters~\eqref{eq:rhotx}.

Before we continue with finding the \pdf\ for multiple swimmers, let us
investigate how the variance of displacements evolves.  The second moment
of~$\Rv_\pal^1$ is
\begin{equation}
  \avg{(\Rc_\pal^1)^2} =
  \int_{\Vol}\rc^2\,\probone_{\Rv_\pal^1}(\rv)\dint\Vol_{\rv}
  =
  \int_\Vol \Delta_\pal^2(\etav)
  \,\frac{\!\dint\Vol_{\etav}}{\Vol}.
\end{equation}
This typically goes to zero as~$\Vol\rightarrow\infty$, since a single swimmer
in an infinite volume shouldn't give any fluctuations on average.  We
write~$\Rv_\pal^\Nswim$ for the random particle displacement due to~$\Nswim$
swimmers; the second moment of~$\Rv_\pal^\Nswim$ is
\begin{equation}
  \avg{(\Rc_\pal^\Nswim)^2} 
  = \Nswim\avg{(\Rc_\pal^1)^2} =
  \nd\int_\Vol \Delta_\pal^2(\etav)
  \dint\Vol_{\etav}
  \label{eq:r2N}
\end{equation}
with~$\nd = \Nswim/\Vol$ the number density of swimmers.  This is nonzero (and
might diverge) in the limit~$\Vol\rightarrow\infty$, reflecting the cumulative
effect of multiple swimmers.  Note that this expression is exact, within the
problem assumptions: it doesn't even require~$\Nswim$ to be large.

\begin{figure}
  \begin{center}
  \subfigure[]{
    \includegraphics[height=.25\textheight]{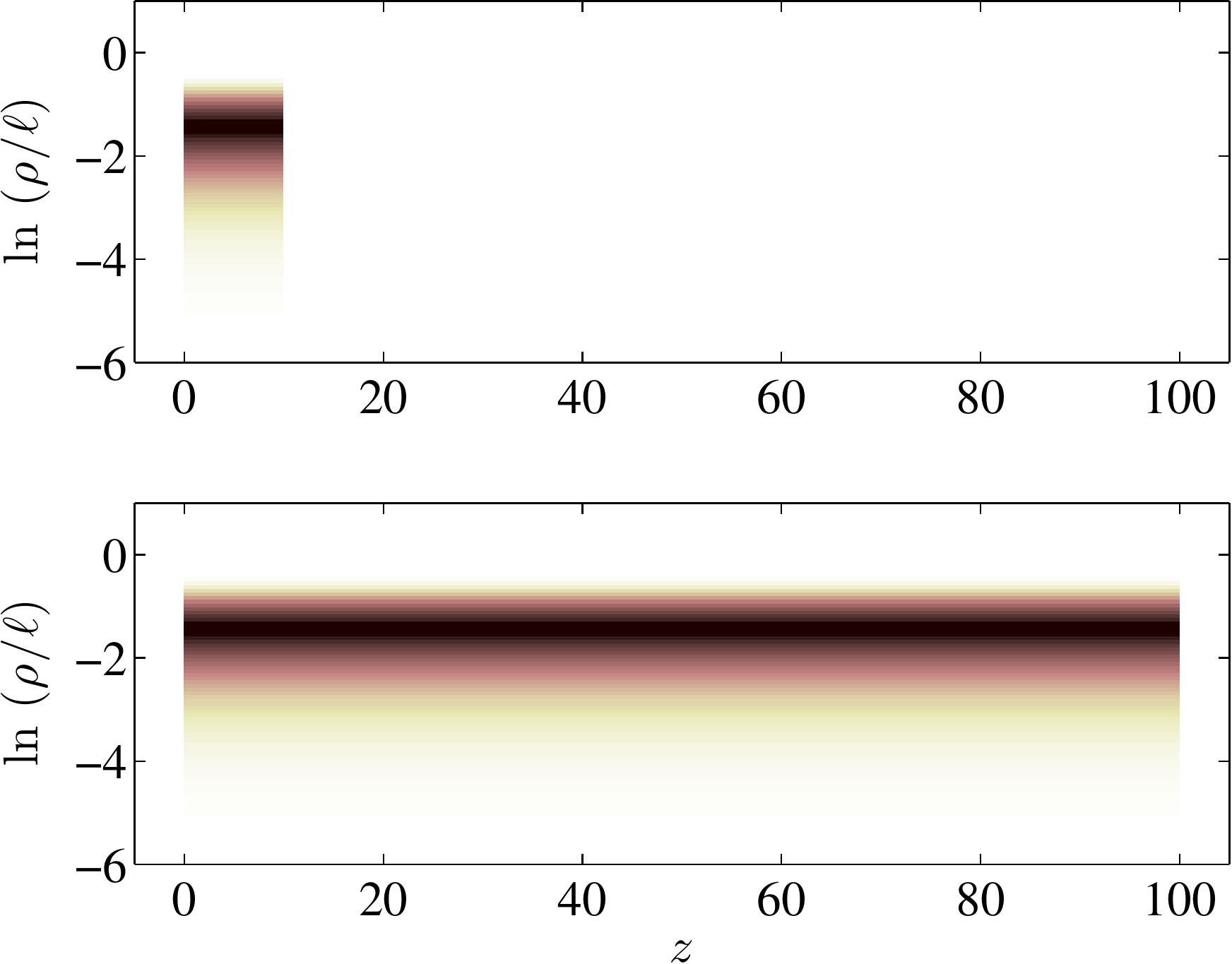}
    \label{fig:sphere_Delta}
  }\hspace{1em}%
  \subfigure[]{
    \includegraphics[height=.25\textheight]{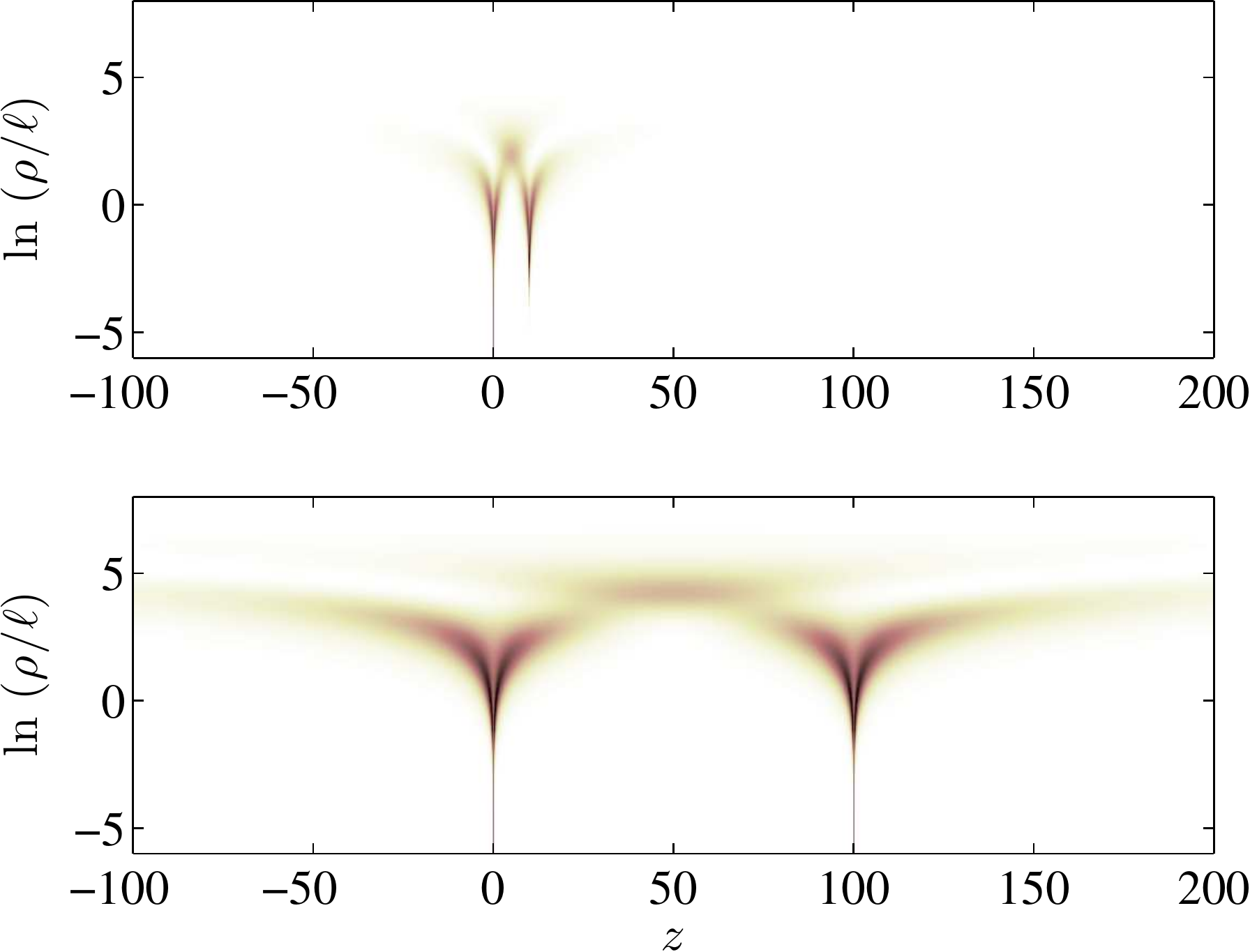}
    \label{fig:stresslet_Delta}
  }
  \end{center}
  \caption{The natural log of~$\rho^2\Delta^2(\rho,\zc)$ (integrand
    of~\eqref{eq:r2N} with~$\dint\Vol_{\etav} =
    2\pi\rho^2\dint(\ln\rho)\dint\zc$) for (a) a sphere of radius~$\lsc=1$ in
    inviscid flow, moving a path length~$\pal=10$ (top) and~$100$ (bottom),
    plotted on the same scale.  The scale of the integrand doesn't change,
    only its support.  Here~$\etav=(\rho,\zc)$ with~$\zc$ the swimming
    direction and~$\rho$ the distance from the~$\zc$ axis.  (b) Same as (a)
    but for a stresslet velocity field.  The integral~\eqref{eq:r2N} grows
    linearly with~$\pal$ for both (a) and (b).}
  \label{fig:sphere_stresslet_Delta}
\end{figure}

The expression~\eqref{eq:r2N} will lead to diffusive behavior if the integral
grows linearly in~$\pal$ (or if the swimmers change direction~\cite{Lin2011},
which we shall not treat here).  Surprisingly, it has been found to do so in
two distinct ways.  In the first, exemplified by bodies in inviscid
flow~\cite{Thiffeault2010b,Lin2011}, the support of~$\Delta_\pal$ grows
linearly with~$\pal$, but the displacements themselves become independent
of~$\pal$ when~$\pal$ is large.  The intuition is that the swimmer pushes
particles a finite distance as it encounters them.  As we wait longer, the
volume of such displaced particles grows linearly in~$\pal$, but once
particles are displaced they are left behind and suffer no further
displacement.  This diffusive behavior is thus appropriate for very localized
interactions, where the only displaced particles are very near the axis of
swimming.  This tends to occur in inviscid flow, or for spherical
`treadmillers' in viscous flow.  See Fig.~\ref{fig:sphere_Delta} for an
illustration.

The second situation in which~\eqref{eq:r2N} shows diffusive behavior even
for straight swimming paths is when the far-field velocity has the form of a
stresslet, as is the case for a force-free swimmer in a viscous fluid.  This
diffusive behavior was observed in \citet{Lin2011} but it was
\citet{Pushkin2013b} who provided a full explanation.  For a stresslet
swimmer, the main contributions to~\eqref{eq:r2N} come
from~$\lVert\etav\rVert$ of order $\pal$, so it is appropriate to use a point
singularity model swimmer for large~$\pal$.  In that case the drift function
has the scaling~$\Delta_\pal(\etav) = \Delta_\pal(\pal\zetav) =
\pal^{-1}\Deltasc(\zetav)$, where~$\zetav=\etav/\pal$ is a dimensionless
variable and the function~$\Deltasc(\zetav)$ is independent of~$\pal$ for
large~$\pal$~\cite{Pushkin2013b}.  Inserting this form in~\eqref{eq:r2N}, we
find
\begin{equation}
  \int \Delta_\pal^2(\etav)\dint\Vol_{\etav}
  =
  \int \l(\pal^{-2}\Deltasc^2(\zetav)\r)\l(\pal^3\dint\Vol_{\zetav}\r)
  \sim \pal.
  \label{eq:Deltasc2int}
\end{equation}
The integral of~$\Deltasc^2(\zetav)$ converges despite having
singularities~\cite{Pushkin2013b}.  We thus see that the integral
in~\eqref{eq:r2N} grows linearly in~$\pal$ for very different reasons than our
first case: here the volume of particles affected by the swimmer grows
as~$\pal^3$ (particles are affected further and further away), but they are
displaced less (since they are further away, see
Fig.~\ref{fig:stresslet_Delta}).  Any truncation of the integral
in~\eqref{eq:Deltasc2int} (because of finite volume effect) will lead to a
decrease in the diffusivity, a possible origin for the decrease in diffusivity
with path length observed in \citet{Jepson2013}.  Note also that the
reorientation mechanism discussed by~\citet{Lin2011} is not necessary in this
case to achieve the diffusive behavior, as pointed out by \citet{Pushkin2014}.

Having addressed the growth of the variance, we continue with finding the
\pdf\ of displacements for multiple swimmers.  We write~$\X_\pal^\Nswim$ for a
single coordinate of~$\Rv_\pal^\Nswim$ (which coordinate is immaterial,
because of isotropy).  From~\eqref{eq:rhotx} with~$\sdim=2$ we can
compute~$\probone_{\X_\pal^1}(\xc)$, the marginal distribution for one
coordinate:
\begin{equation}
  \probone_{\X_\pal^1}(\xc)
  = \int_{-\infty}^\infty\probone_{\Rv_\pal^1}(\rv)\dint\yc
  = \int_\Vol\int_{-\infty}^\infty\frac{1}{2\pi\rc}\,
  \delta(\rc - \Delta_\pal(\etav))
  \dint\yc\,\frac{\!\dint\Vol_{\etav}}{\Vol}.
\end{equation}
Since~$\rc^2=\xc^2+\yc^2$, the~$\delta$-function will capture two values
of~$\yc$, and with the Jacobian included we obtain
\begin{equation}
  \probone_{\X_\pal^1}(\xc)
  = \frac{1}{\pi}\int_\Vol
  \frac{1}{\sqrt{\Delta_\pal^2(\etav) - \xc^2}}
  \indf{\Delta_\pal(\etav) > \lvert\xc\rvert}
  \,\frac{\!\dint\Vol_{\etav}}{\Vol}\,,
  \label{eq:rhotxc2D}
\end{equation}
where~$\indf{A}$ is an indicator function: it is~$1$ if~$A$ is satisfied, $0$
otherwise.

The marginal distribution in the three-dimensional case proceeds the same way
from~\eqref{eq:rhotx} with~$\sdim=3$:
\begin{equation}
  \probone_{\X_\pal^1}(\xc)
  = \int_{-\infty}^\infty\probone_{\Rv_\pal^1}(\rv)\dint\yc\dint\zc
  = \int_\Vol\int_{-\infty}^\infty\int_{-\infty}^\infty\frac{1}{4\pi\rc^2}\,
  \delta(\rc - \Delta_\pal(\etav))
  \dint\yc\dint\zc\,\frac{\!\dint\Vol_{\etav}}{\Vol}.
\end{equation}
Again with~$\rc^2=\xc^2+\yc^2+\zc^2$ the~$\delta$-function captures two
values of~$\zc$, and with the Jacobian included we obtain
\begin{equation}
  \probone_{\X_\pal^1}(\xc)
  = \frac{1}{2\pi}\int_\Vol\int_{-\infty}^\infty
  \frac{1}{\Delta_\pal(\etav)}\,
  \frac{1}{\sqrt{\Delta_\pal^2(\etav) - \xc^2 - \yc^2}}
  \indf{\Delta_\pal^2(\etav) > \xc^2+\yc^2}
  \!\dint\yc\,\frac{\!\dint\Vol_{\etav}}{\Vol}.
\end{equation}
Now we integrate over~$\yc$ to get
\begin{equation}
  \probone_{\X_\pal^1}(\xc)
  = \tfrac{1}{2}\int_\Vol
  \frac{1}{\Delta_\pal(\etav)}\,
  \indf{\Delta_\pal(\etav) > \lvert\xc\rvert}
  \,\frac{\!\dint\Vol_{\etav}}{\Vol}
  \label{eq:rhotxc3D}
\end{equation}
which is the three-dimensional analogue of~\eqref{eq:rhotxc2D}.  The integrand
of~\eqref{eq:rhotxc3D} has an intuitive interpretation.  The indicator
function says that a displacement in a random direction must at least be
larger than~$\lvert\xc\rvert$ to project to a value~$\xc$.  The factor
of~$\Delta_\pal(\etav)$ in the denominator then tells us that large
displacements in a random direction are less likely to project to a
value~$\xc$.  (The two-dimensional form~\eqref{eq:rhotxc2D} has essentially
the same interpretation, with a different weight.)

In order to sum the displacements due to multiple swimmers, we need the
characteristic function of~$\probone_{\X_\pal^1}(\xc)$, defined by
\begin{equation}
  \avg{\ee^{\imi\kc\X_\pal^1}} = \int_{-\infty}^\infty\probone_{\X_\pal^1}(\xc)\,
  \ee^{\imi\kc\xc}\dint\xc.
\end{equation}
For the two-dimensional \pdf~\eqref{eq:rhotxc2D}, we have
\begin{equation}
  \avg{\ee^{\imi\kc\X_\pal^1}}
  =
  \int_\Vol J_0(\kc\Delta_\pal(\etav))
  \,\frac{\!\dint\Vol_{\etav}}{\Vol}
  \label{eq:char2}
\end{equation}
where~$J_0(x)$ is a Bessel function of the first kind.
For the three-dimensional \pdf~\eqref{eq:rhotxc3D}, the characteristic
function is
\begin{equation}
  \avg{\ee^{\imi\kc\X_\pal^1}}
  =
  \int_\Vol
  \sinc{(\kc\Delta_\pal(\etav))}\,\frac{\!\dint\Vol_{\etav}}{\Vol}
  \label{eq:char3}
\end{equation}
where~$\sinc \xc \ldef \xc^{-1}\sin \xc$ for~$x\ne 0$, and~$\sinc 0 \ldef
1$.\footnote{Beware that this function is sometimes defined as~$(\pi
  \xc)^{-1}\sin (\pi \xc)$, most notably by Matlab.}  The
expression~\eqref{eq:char3} appears in~\cite{Pushkin2014}, except here we
compute it directly from a spatial integral rather than from the \pdf\ of
$\Delta$.  The main difference will come in the way we take the limit $\Vol
\rightarrow \infty$ below, which will allow us to study the number density
dependence directly.

We define
\begin{equation}
  \K(\xc) \ldef \begin{cases}
     1 - J_0(\xc),\quad &\sdim=2;\\
     1 - \sinc\xc,\quad &\sdim=3,
   \end{cases}
  \label{eq:Ksdimdef}
\end{equation}
We have~$\K(0)=\K'(0)=0$, $\K''(0)=1/\sdim$,
so~$\K(\xi) \sim (1/2\sdim)\,\xi^2 + \Order{\xi^4}$
as~$\xi\rightarrow0$.  For large argument, $\K(\xi)\rightarrow 1$.
We can then write the two cases~\eqref{eq:char2}--\eqref{eq:char3} for the
characteristic function together as
\begin{equation}
  \avg{\ee^{\imi\kc\X_\pal^1}}
  =
  1 - (\vpal/\Vol)\,\cK_\pal(\kc)
  \label{eq:charf}
\end{equation}
where
\begin{equation}
  \cK_\pal(\kc) \ldef
  \frac{1}{\vpal}
  \int_\Vol\K(\kc\Delta_\pal(\etav))\dint\Vol_{\etav}.
  \label{eq:cKdef}
\end{equation}
Here~$\vpal$ is the volume `carved out' by a swimmer moving a distance~$\pal$:
\begin{equation}
  \vpal = \pal\areasw
  \label{eq:vpal}
\end{equation}
with~$\areasw$ the cross-sectional area of the swimmer in the direction of
motion.

Since we are summing independent particle displacements, the probability
distribution of the sum is the convolution of~$\Nswim$ one-swimmer
distributions.  Using the Fourier transform convolution property, the
characteristic function for~$\Nswim$ swimmers is
thus~$\avg{\ee^{\imi\kc\X_\pal^\Nswim}} =
\avg{\ee^{\imi\kc\X_\pal^1}}^\Nswim$.  From~\eqref{eq:charf},
\begin{equation}
  \avg{\ee^{\imi\kc\X_\pal^1}}^\Nswim
  =
  \l(1 - \vpal\cK_\pal(\kc)/\Vol\r)^{\nd\Vol},
  \label{eq:charfNswim}
\end{equation}
where we used~$\Nswim = \nd\Vol$, with~$\nd$ the number density of swimmers.
We will need the following simple result:
\begin{prop}
\label{prop:expid}
Let~$\yprop(\eprop) \sim \order{\eprop^{-\Mprop/(\Mprop+1)}}$
as~$\eprop\rightarrow 0$ for an integer~$\Mprop \ge 1$; then
\begin{equation}
  (1 - \eprop\yprop(\eprop))^{1/\eprop}
  = \exp\biggl(-\sum_{m=1}^{\Mprop}\frac{\eprop^{m-1}\yprop^m(\eprop)}{m}\biggr)
  \l(1 + \order{\eprop^0}\r), \quad \eprop \rightarrow 0.
  \label{eq:expid}
\end{equation}
\end{prop}
\noindent
See Appendix~\ref{apx:proof} for a short proof.

Let's examine the assumption of Proposition~\ref{prop:expid} for~$\Mprop=1$
applied to~\eqref{eq:charfNswim}, with~$\eprop=1/\Vol$
and~$\yprop=\vpal\cK_\pal(\kc)$.
For~$\Mprop=1$, the assumption of Proposition~\ref{prop:expid} requires
\begin{equation}
  \cK_\pal(\kc) \sim \order{\Vol^{1/2}},\qquad\Vol\rightarrow\infty.
  \label{eq:propcond2}
\end{equation}
A stronger divergence with~$\Vol$ means using a larger~$\Mprop$ in
Proposition~\ref{prop:expid}, but we shall not need to consider this here.
Note that it is not possible for~$\cK_\pal(\kc)$ to diverge faster
than~$\Order{\Vol}$, since~$\gamma(\xc)$ is bounded.  In order
for~$\cK_\pal(\kc)$ to diverge as~$\Order{\Vol}$, the displacement must be
nonzero as~$\Vol\rightarrow\infty$, an unlikely situation that can be ruled
out.

Assuming that~\eqref{eq:propcond2} is satisfied, we use
Proposition~\ref{prop:expid} with~$\Mprop=1$ to make the large-volume
approximation
\begin{equation}
  \avg{\ee^{\imi\kc\X_\pal^1}}^\Nswim
  =
  \l(1 - \vpal\cK_\pal(\kc)/\Vol\r)^{\nd\Vol}
  \sim
  \exp\l(-\nd\vpal\,\cK_\pal(\kc)\r), \quad
  \Vol \rightarrow \infty.
  \label{eq:largevol}
\end{equation}
If the integral~$\cK_\pal(\kc)$ is convergent as~$\Vol\rightarrow\infty$ we
have achieved a volume-independent form for the characteristic function, and
hence for the distribution of~$\xc$ for a fixed swimmer density.  We define
the quantity
\begin{equation}
  \Mpal \ldef \nd\vpal = \pal/\mfp
  \label{eq:Mpal}
\end{equation}
where~$\mfp = (\nd\areasw)^{-1}$ is the swimmer mean free path.  Since $\vpal$
is the volume carved out by a single swimmer moving a distance~$\pal$
(Eq.~\eqref{eq:vpal}), $\Mpal$ is the expected number of swimmers that will
`hit' a given fluid particle.

A comment is in order about evaluating~\eqref{eq:cKdef} numerically: if we
take~$\lvert\kc\rvert$ to~$\infty$, then~$\K(\kc\Delta) \rightarrow 1$, and
thus $\vpal\cK \rightarrow \Vol$, which then leads to~$\ee^{-\Nswim}$
in~\eqref{eq:largevol}.  This is negligible as long as the number of
swimmers~$\Nswim$ is moderately large.  In practice, this means
that~$\lvert\kc\rvert$ only needs to be large enough that the argument of the
decaying exponential in~\eqref{eq:largevol} is of order one, that is
\begin{equation}
  \Mpal\,\cK_\pal(\kmax) \sim \Order{1}.
  \label{eq:kmax}
\end{equation}
Wavenumbers~$\lvert\kc\rvert > \kmax$ do not contribute
to~\eqref{eq:largevol}.  (We are assuming monotonicity of~$\cK_\pal(\kc)$
for~$\kc>0$, which will hold for our case.)  Note that~\eqref{eq:kmax} implies
that we need larger wavenumbers for smaller densities~$\nd$: a typical fluid
particle then encounters very few swimmers, and the distribution should be far
from Gaussian.

Now that we've computed the characteristic function for~$\Nswim$
swimmers~\eqref{eq:largevol}, we finally recover the \pdf\ of~$\xc$
for~$\Nswim=\nd\Vol$ swimmers as the inverse Fourier transform
\begin{equation}
  \prob_{\X_\pal}(\xc) = \frac{1}{2\pi}\int_{-\infty}^\infty
  \exp\l(-\Mpal\,\cK_\pal(\kc)\r)
  \ee^{-\imi\kc\xc}\dint\kc,
  \label{eq:rhotxNswim}
\end{equation}
where we dropped the superscript~$\Nswim$ from~$\X_\pal^\Nswim$ since the
number of swimmers no longer enters the expression directly.

\section{Comparing to experiments}
\label{sec:numerics}

We now compare the theory discussed in the previous sections to the
experiments of \citeauthoreos{Leptos2009}, in particular the observed
dependence of the distribution on the number density~$\phi$.  (Another aspect
of their experiments, the `diffusive scaling' of the distributions, will be
discussed in Section~\ref{sec:diffscal}.)  In their experiments they use the
microorganism \textit{C.\ reinhardtii}, an alga of the `puller' type, since
its two flagella are frontal.  This organism has a roughly spherical body with
radius~$\lsc \approx 5\,\microm$.  They observe a distribution of swimming
speeds with a strong peak around~$100\,\microm/\second$.  They place
fluorescent microspheres of about a micron in radius in the fluid, and
optically measure their displacement as the organisms move.  The volume
fraction of organisms varies from~$\phi=0\%$ (pure fluid) to~$2.2\%$.

They measure the displacement of the microspheres along a reference direction,
arbitrarily called~$\xc$ (the system is assumed isotropic).  Observing many
microspheres allows them to compute the \pdf\ of tracer
displacements~$\X_\pal$, which we've denoted~$\prob_{\X_\pal}(\xc)$.  Thus,
$\prob_{\X_\pal}(\xc)\dint\xc$ is the probability of observing a particle
displacement~$\X_\pal \in [\xc,\xc+\dint\xc]$ after a path length~$\pal$.
(They write their density~$P(\Delta x,\Delta t)$, where~$(\Delta x,\Delta t)$
are the same as our~$(\xc,\pal/\Uc)$.)

At zero volume fraction ($\phi=0$), the \pdf\ $\prob_{\X_\pal}(\xc)$ is
Gaussian, due solely to thermal noise.  For higher number densities,
\citeauthor{Leptos2009} see exponential tails appear and the Gaussian core
broaden.  The distribution is well-fitted by the sum of a Gaussian and an
exponential:
\begin{equation}
  \prob_{\X_\pal}(\xc) = \frac{1-f}{\sqrt{2\pi\delta_{\text{g}}^2}}
  \,\ee^{-\xc^2/2\delta_{\text{g}}^2}
  + \frac{f}{2\delta_{\text{e}}}\,\ee^{-\lvert\xc\rvert/\delta_{\text{e}}}.
  \label{eq:nonGaussform}
\end{equation}
They observe the scalings~$\delta_{\text{g}} \approx A_{\text{g}}\tt^{1/2}$
and~$\delta_{\text{e}} \approx A_{\text{e}}\tt^{1/2}$, where~$A_{\text{g}}$
and~$A_{\text{e}}$ depend on~$\phi$.  The dependence on~$\tt^{1/2}$ is
referred to as the `diffusive scaling' and will be discussed in
Section~\ref{sec:diffscal}.  Exploiting this scaling, \citet{Eckhardt2012}
have fitted these distributions very well to a continuous-time random walk
model, but this does not suggest a mechanism.

\begin{figure}
  \begin{center}
    \includegraphics[width=.7\textwidth]{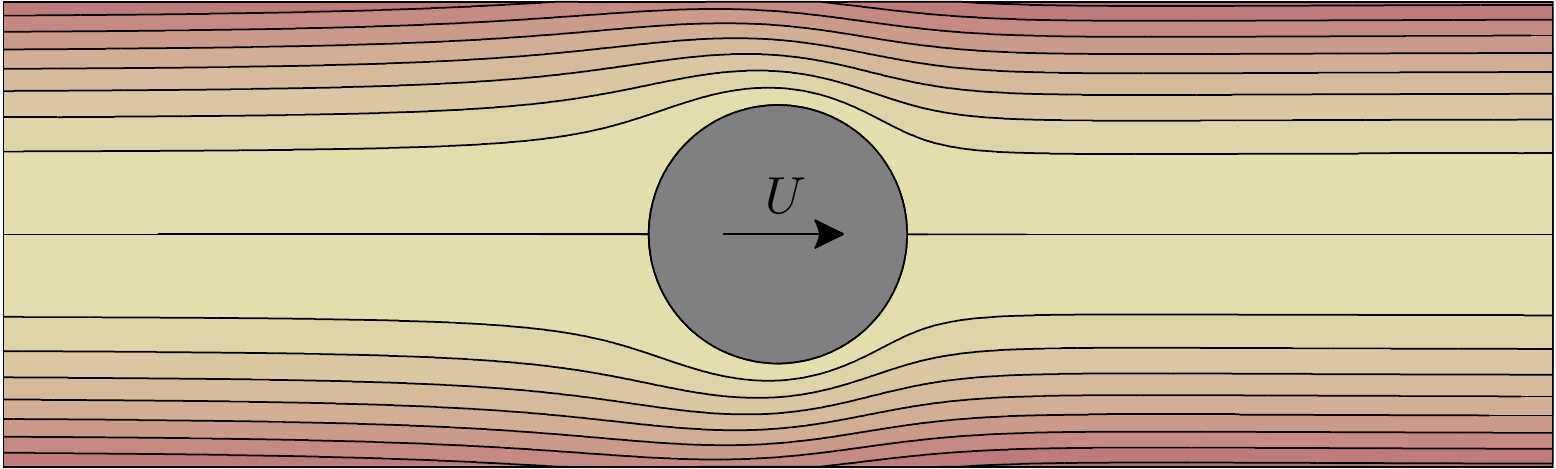}
  \end{center}
  \caption{Contour lines for the axisymmetric streamfunction of a squirmer of
    the form~\eqref{eq:squirm_strfcn}, with~\hbox{$\beta=0.5$}.  This swimmer
    is of the puller type, as for \textit{C.\ reinhardtii}.}
  \label{fig:squirmer_contour_beta=0p5}
\end{figure}

We shall use a model swimmer of
the squirmer type~\cite{Lighthill1952,Blake1971,Ishikawa2006, Ishikawa2007b,
  Drescher2009}, with axisymmetric streamfunction~\cite{Lin2011}
\begin{equation}
  \Psi_{\text{sf}}(\rho,z)
  = \tfrac12{\rho^2\,\Uc}\l\{-1 + \frac{\lsc^3}{(\rho^2+z^2)^{3/2}}
  + \tfrac32\frac{\beta \lsc^2 z}{(\rho^2+z^2)^{3/2}}
  \l(\frac{\lsc^2}{\rho^2+z^2} - 1\r)\r\}
  \label{eq:squirm_strfcn}
\end{equation}
in a frame moving at speed~$\Uc$.  Here~$z$ is the swimming direction
and~$\rho$ is the distance from the~$z$ axis.  To mimic \textit{C.\
  reinhardtii}, we use~$\lsc=5\,\microm$ and $\Uc=100\,\microm/\second$.
(\citet{Leptos2009} observe a distribution of velocities but the peak is
near~$100\,\microm/\second$.)  We take~$\beta=0.5$ for the relative stresslet
strength, which gives a swimmer of the puller type, just like \textit{C.\
  reinhardtii}.  The contour lines of the axisymmetric
streamfunction~\eqref{eq:squirm_strfcn} are depicted in
Fig.~\ref{fig:squirmer_contour_beta=0p5}.  The parameter~$\beta=0.5$ is the
only one that was fitted (visually) to give good agreement later.

First we compute the drift function~$\Delta_\pal(\etav)$ for a single swimmer
moving along the~$\zc$ axis.  The model swimmer is axially symmetric,
so~$\etav$ can be written in terms of~$\zc$ and~$\rho$, the perpendicular
distance to the swimming axis.  We take~$\pal=12\,\microm$, since the time
is~$\tt=0.12\,\second$ in Fig.~2(a) of \citeauthor{Leptos2009}, and our
swimmer moves at speed~$\Uc=100\,\microm/\second$.  We
compute~$\Delta_\pal(\rho,\zc)$ for a large grid of~$\ln\rho$ and~$\zc$
values, using the analytic far-field stresslet form for the
displacement~\cite{Dunkel2010,Mino2013,Pushkin2013b} when far away from the
swimmer's path.

From the drift function~$\Delta_\pal(\etav)$ we now want to
compute~$\cK_\pal(\kc)$ defined by~\eqref{eq:cKdef}.  To estimate how large
a~$\kc$ value we will need, we start from the smallest volume fraction in the
experiments, $\phi \sim 0.4\%$.  For spherical swimmers of radius~$\lsc \sim
5\,\microm$ (with cross-sectional area~$\areasw = \pi\lsc^2 \sim
78.5\,\microm^2$), this gives a number density of~$7.6 \times
10^{-6}\,\microm^{-3}$.  We thus get~$\Mpal = \nd\areasw\pal \sim 7.2\times
10^{-3}$.  The criterion~\eqref{eq:kmax} then tells us that we need~$\kmax$
large enough that $\cK_\pal(\kmax) \sim 1/\Mpal \sim 139$.
\begin{figure}
  \begin{center}
    \includegraphics[width=.5\textwidth]{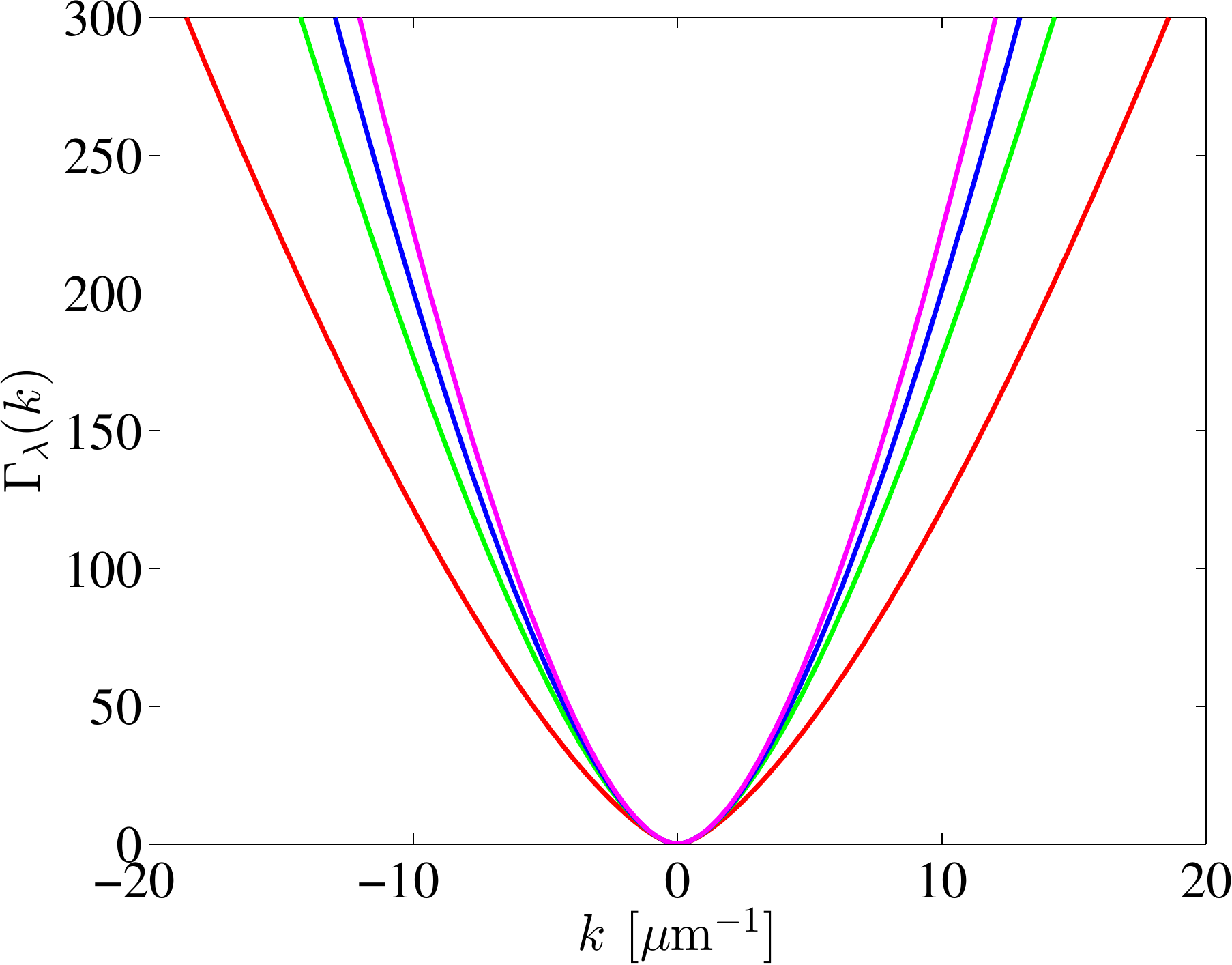}
  \end{center}
  \caption{\protect\coloronline The function $\cK_\pal(\kc)$ defined by
    Eq.~\eqref{eq:cKdef} for (from broadest to narrowest)~$\pal=12\,\microm$,
    $36\,\microm$, $60\,\microm$, and $96\,\microm$.}
  \label{fig:pdfX1_charfun}
\end{figure}
Figure~\ref{fig:pdfX1_charfun} shows the numerically-computed~$\cK_\pal(\kc)$
for several values of~$\pal$, with~$\pal=12\,\microm$ the broadest curve.  We
can see from the figure that choosing~$\kmax \sim 20\,\microm^{-1}$ will
ensure that~$\Mpal\cK_\pal(\kmax)$ is large enough.  As~$\pal$ gets larger,
$\kmax$ decreases, reflecting the trend towards the central limit theorem
(which corresponds to the small-$\kc$ expansion of~$\cK_\pal(\kc)$, see
Section~\ref{sec:largedev}).  Note also that~$\cK_\pal(\kc)$ tends to become
independent of~$\pal$ as~$\pal$ gets larger.

To obtain~$\prob_{\X_\pal}(\xc)$ and compare to \citeauthor{Leptos2009}, we
must now take the inverse Fourier transform of~$\exp(-\Mpal\cK_\pal(\kc))$, as
dictated by~\eqref{eq:rhotxNswim}.  This is straightforward using Matlab's
\texttt{ifft} routine.  The `period' (domain in~$\xc$) is controlled by the
spacing of the $\kc$ grid, so we make sure the grid is fine enough to give us
the largest values of~$\xc$ required.  We also convolve with a Gaussian
distribution of half-width~$\sqrt{2\Diff_0\tt}=0.26\,\microm$ to mimic thermal
noise.  This follows from the value~$\Diff_0=0.28\,\microm^2/\second$ measured
by \citeauthor{Leptos2009} for the diffusivity of the microspheres.  The value
of~$\Diff_0$ is consistent with the Stokes--Einstein equation for the
diffusivity of thermally-agitated small spheres in a fluid.

\begin{figure}
  \begin{center}
  \subfigure[]{
    \raisebox{.1em}{
    \includegraphics[width=.49\textwidth]{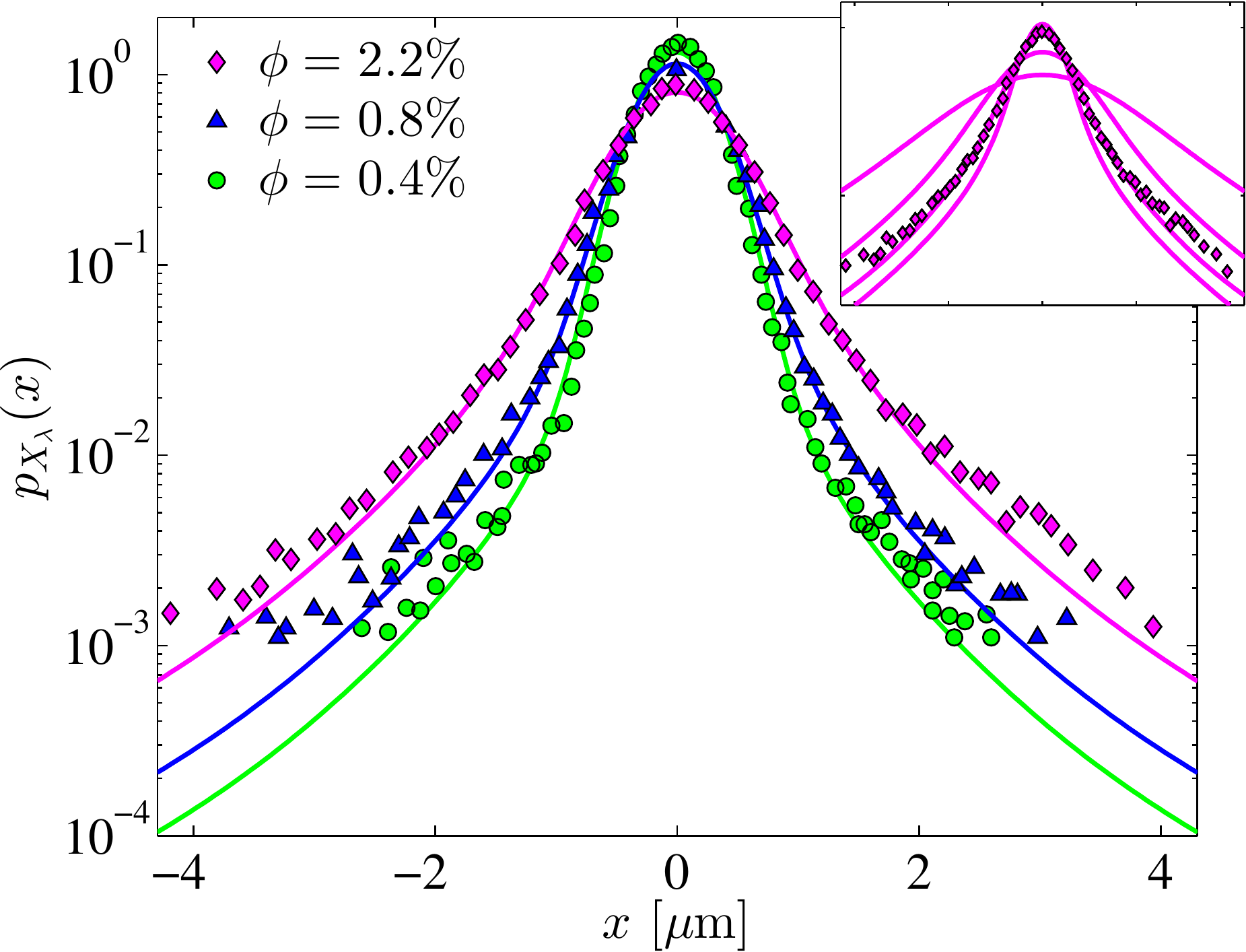}}
    \label{fig:compare_to_Leptos}
  }%
  \subfigure[]{
    \includegraphics[width=.477\textwidth]{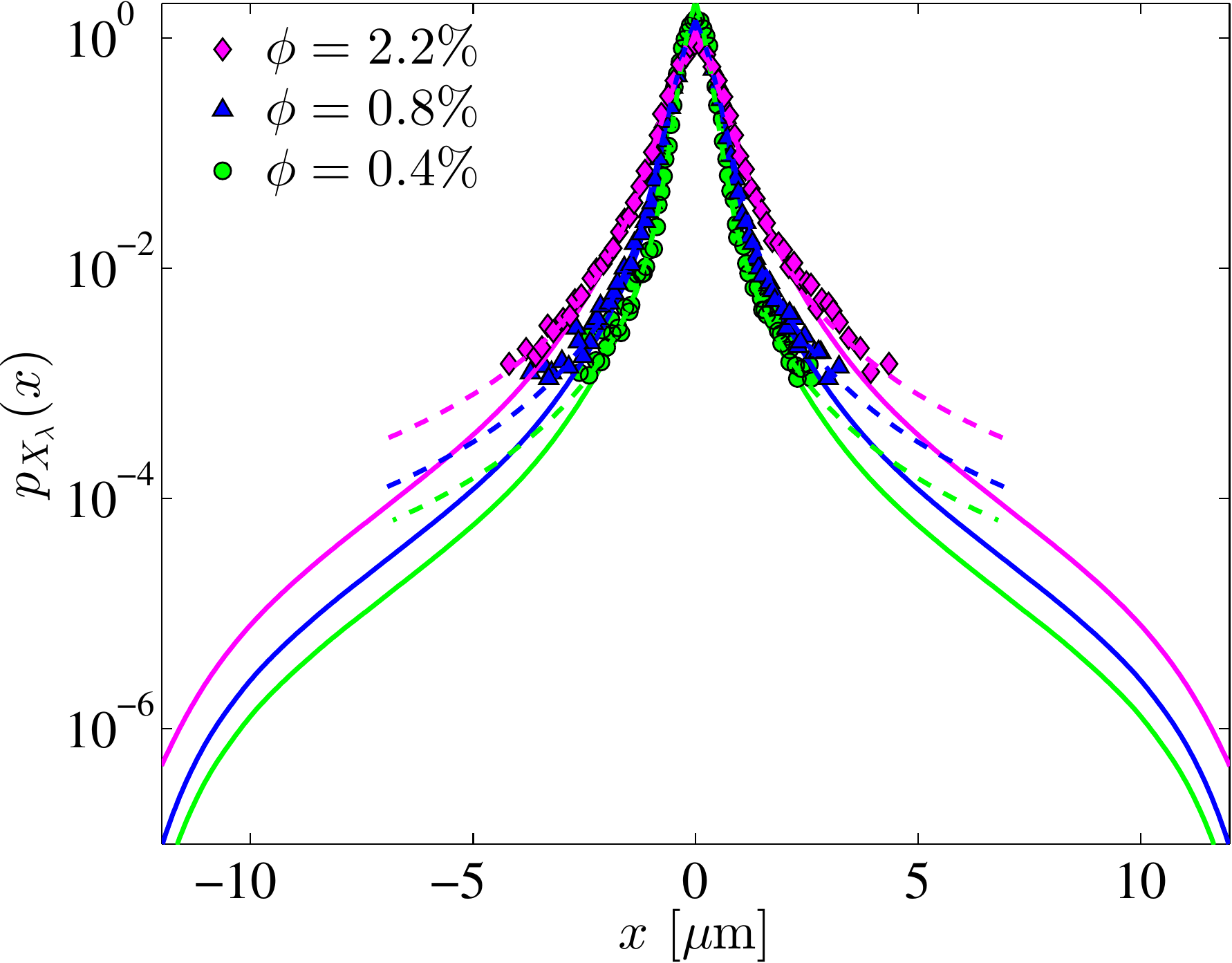}
    \label{fig:compare_to_Eckhardt}
  }
  \end{center}
  \caption{\protect\coloronline (a) The \pdf\ of particle displacements after
    a path length~$\pal=12\,\microm$, for several values of the volume
    fraction~$\phi$.  The data is from \citet{Leptos2009}, and the figure
    should be compared to their Fig.~2(a).  The theoretical curves were
    obtained from~\eqref{eq:rhotxNswim} for the model squirmer in
    Fig.~\ref{fig:squirmer_contour_beta=0p5}, with some noise corresponding to
    thermal diffusivity as measured in \citet{Leptos2009}.  Inset: comparison
    of (from broadest to narrowest)~$\beta = 2$, $1$, $0.5$, and~$0.1$,
    for~$\phi=2.2\%$, showing the sensitivity of the fit~$\beta=0.5$.  (b)
    Same as (a) but on a wider scale, also showing the form suggested
    by~\citet{Eckhardt2012} (dashed lines).}
  \label{fig:compare_to_Leptos_both}
\end{figure}

The results are plotted in Fig.~\ref{fig:compare_to_Leptos} and compared to
the data of Fig.~2(a) of \citet{Leptos2009}.  The agreement is excellent: we
remind the reader that we adjusted only one parameter, $\beta=0.5$.  This
parameter was visually adjusted to the~$\phi=2.2\%$ data in
Fig.~\ref{fig:compare_to_Leptos}, since the larger concentration is most
sensitive to~$\beta$; a more careful fit is unnecessary given the
uncertainties in both model and data.  (The inset shows the sensitivity of the
fit to~$\beta$.)  All the other physical quantities were gleaned from
\citeauthoreos{Leptos2009}.  What is most remarkable about the agreement in
Fig.~\ref{fig:compare_to_Leptos} is that it was obtained using a model
swimmer, the spherical squirmer, which is not expected to be such a good model
for \textit{C.\ reinhardtii}.  The real organisms are strongly time-dependent,
for instance, and do not move in a perfect straight line.  Nevertheless the
model captures very well the \pdf\ of displacements, in particular the volume
fraction dependence.  The model swimmer slightly underpredicts the tails, but
since the tails are associated to large displacements they depend on the
near-field details of the swimmer, so it is not surprising that our model
swimmer should deviate from the data.

In Figure~\ref{fig:compare_to_Eckhardt} we compare our results to the
phenomenological fit of \citet{Eckhardt2012} based on continuous-time random
walks: their fit is better in the tails, but our models disagree immediately
after the data runs out.  Our model has the realistic feature that the
distribution is cut off at the path length~$\pal = 12\,\microm$, since it is
extremely unlikely that a particle had two close encounters with a swimmer at
these low volume fractions.

\begin{figure}
  \begin{center}
    \includegraphics[width=.6\textwidth]{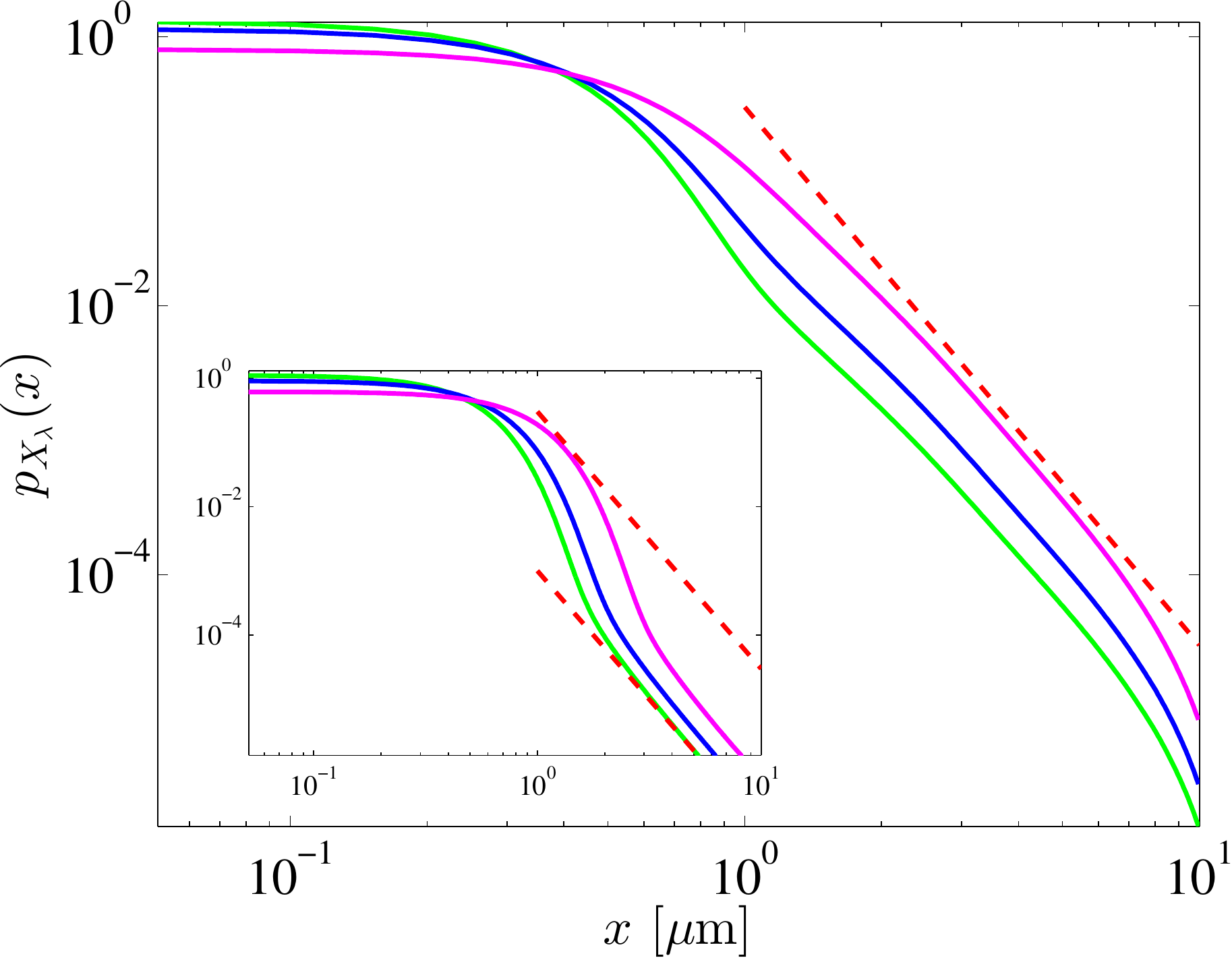}
  \end{center}
  \caption{\protect\coloronline The same distributions as in
    Fig.~\ref{fig:compare_to_Leptos}, but on a log-log plot.  The dashed line
    is the $\xc^{-4}$ power law predicted by \citet{Pushkin2014}.  Inset:
    numerical simulation with only the stresslet far-field displacement
    included.}
  \label{fig:compare_to_PY}
\end{figure}

A possible explanation as to why the squirmer model does so well was provided
by \citet{Pushkin2014}.  They used numerical simulations of squirmers (with a
larger value~$\beta=2$ that leads to a trapped volume) to show that the tails
of distribution scale as~$\xc^{-4}$, which is the asymptotic form of the
stresslet displacement distribution.  Figure~\ref{fig:compare_to_PY} shows
that our computations have a similar tail, though we emphasize here that our
agreement with the experiments of \citet{Leptos2009} is \emph{quantitative}
and correctly reproduces the volume fraction dependence.  We also point out
that though the trend in Fig.~\ref{fig:compare_to_PY} follows~$\xc^{-4}$, the
slope changes gradually and does not have a clear power law (the log scale
means the deviations are quite large).  The inset in
Fig.~\ref{fig:compare_to_PY} is a numerical simulation that includes only the
singularity in the stresslet displacement, $\Delta(\etav) \sim
\lVert\etav\rVert^{-1}$, as assumed in the analysis of \citet{Pushkin2014}.
Though the~$\xc^{-4}$ tails are eventually achieved, they have far lower
probability than needed to explain the numerics.  \citeauthor{Pushkin2014}'s
use of the far-field stresslet form to predict the tails is thus questionable,
at least for short path lengths.

\begin{figure}
  \begin{center}
  \subfigure[]{
    \includegraphics[height=.25\textheight]{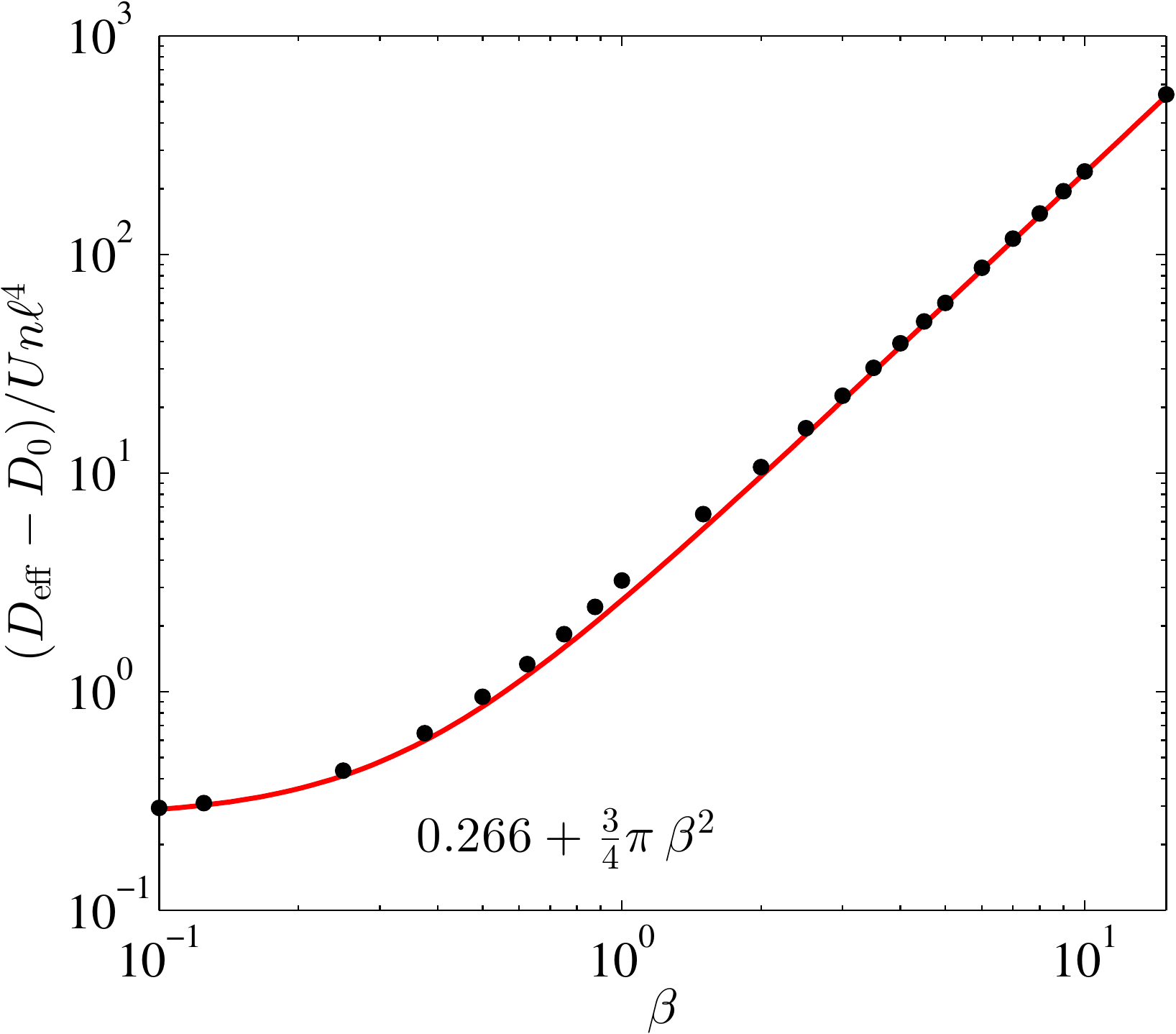}
    \label{fig:squirmer_Deff_vs_beta}
  }\hspace{1em}%
  \subfigure[]{
    \includegraphics[height=.247\textheight]{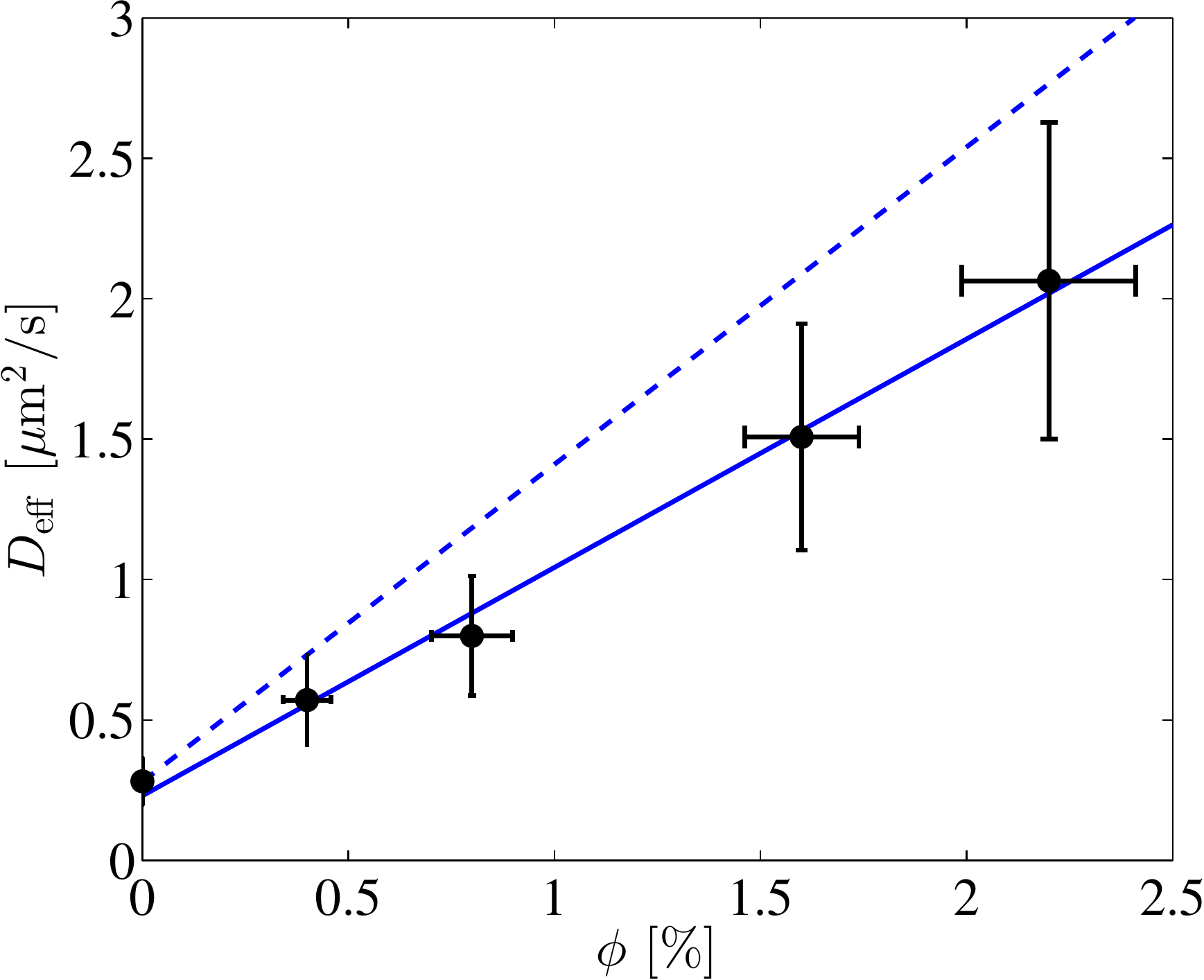}
    \label{fig:compare_to_Leptos_fig3b}
  }
  \end{center}
  \caption{(a) For the squirmer model~\eqref{eq:squirm_strfcn}, dependence of
    the effective diffusivity~$\Diff_{\text{eff}}$ on the stresslet
    strength~$\beta$.  For small~$\beta$, we recover the value for spheres in
    inviscid flow~\cite{Thiffeault2010b}.  An approximate formula is also
    shown as a solid curve.  (b) Comparison of the effective diffusivity data
    from \citet{Leptos2009}, showing their fit (solid line).  The dashed line
    is the prediction for~$\beta=0.5$, used in this paper.}
  \label{fig:squirmer_Deff}
\end{figure}

For the effective diffusivity, \citet{Leptos2009} give the
formula~$\Diff_{\text{eff}} \simeq \Diff_0 + \alpha\,\phi$,
with~$\Diff_0=0.23\,\microm/\second^2$ and~$\alpha = 81.3\,\microm/\second^2$.
Elsewhere in their paper they also give~$\Diff_0=0.28\,\microm/\second^2$ for
the diffusivity of the microspheres in the absence of swimmers, but their
fitting procedure changes the intercept slightly.  (Here we
used~$\Diff_0=0.28\,\microm/\second^2$, but the difference is minute.)
Figure~\ref{fig:squirmer_Deff_vs_beta} shows the numerically-computed
effective diffusivity for our squirmer model, as a function of~$\beta$.  This
curve is as in \cite{Lin2011}, Fig.~6(a), except that we corrected the
integrals in the far field using the analytic expression of
\citet{Pushkin2013b}, which gives a more accurate result.  The Figure also
shows the fit
\begin{equation}
  \frac{\Diff_{\text{eff}} - \Diff_0}{\Uc\nd\lsc^4}
  \simeq 0.266 + \tfrac34\pi\beta^2,
  \label{eq:Deff}
\end{equation}
which is fairly good over the whole range (keeping in mind that this is a
logarithmic plot, so the discrepancy at moderate~$\beta$ are of the order
of~$20$--$30\%$).  Here the value~$0.266$ is the diffusivity due to spheres in
inviscid flow ($\beta=0$, see~\cite{Thiffeault2010b}), and
$\tfrac34\pi\beta^2$ is the large-$\beta$ analytic
expression~\cite{Pushkin2013b} for stresslets.  From the data in
Fig.~\ref{fig:squirmer_Deff_vs_beta} we find~$\alpha \simeq
113\,\microm/\second^2$, significantly larger than \citet{Leptos2009}, as can
be seen in Fig.~\ref{fig:compare_to_Leptos_fig3b}.  The solid line is their
fit, the dashed is our model prediction for~$\beta=0.5$.  The overestimate is
likely due to the method of fitting to the squared displacement: their
Fig.~3(a) clearly shows a change in slope with time, and the early times tend
to be steeper, which would increase the effective diffusivity.  Note also that
their Fig.~3(a) has a much longer temporal range than their PDFs, going all
the way to~$2\,\second$ (compared to~$0.3\,\second$), raising the possibility
that particles were lost by moving out of the focal plane.

\section{The `interaction' viewpoint}
\label{sec:interact}

Equation~\eqref{eq:rhotxNswim} gives the exact solution for the distribution
of uncorrelated displacements due to swimmers of number density~$\nd$.  In
this section we derive an alternative form, in terms of an infinite series,
which is often useful and provides an elegant interpretation
for~\eqref{eq:rhotxNswim}.

The displacement~$\Delta_\pal(\etav)$ typically decays rapidly away from the
swimmer, so that it may often be taken to vanish outside a specified
`interaction volume'~$\Volint$.  Then from~\eqref{eq:cKdef}, since~$\K(0)=0$,
we have
\begin{equation}
  \cK_\pal(\kc)
  = \frac{1}{\vpal}
  \int_{\Volint}\K(\kc\Delta_\pal(\etav))\dint\Vol_{\etav} \\
  =
  \frac{\Volint}{\vpal}\l(1 - \cKm_\pal(\kc)\r)
  \label{eq:cKm0}
\end{equation}
where
\begin{equation}
  \cKm_\pal(\kc) =
  \frac{1}{\Volint}
  \int_{\Volint}(1-\K(\kc\Delta_\pal(\etav)))\dint\Vol_{\etav}\,.
  \label{eq:cKm}
\end{equation}
Define~$\Mpalm \ldef \nd\Volint$; we insert~\eqref{eq:cKm0}
into~\eqref{eq:rhotxNswim} and Taylor expand the exponential to obtain
\begin{equation}
  \prob_{\X_\pal}(\xc) =
  \sum_{m=0}^\infty\frac{\Mpalm^m}{m!}\,\ee^{-\Mpalm}\,
  \frac{1}{2\pi}\int_{-\infty}^\infty
  \cKm_\pal^m(\kc)\,\ee^{-\imi\kc\xc}\dint\kc.
  \label{eq:interac}
\end{equation}
The factor~$\Mpalm^m\,\ee^{-\Mpalm}/m!$ is a Poisson distribution for the
number of `interactions' $m$ between swimmers and a particle: it measures the
probability of finding~$m$ swimmers inside the volume~$\Volint$.  The inverse
transform in~\eqref{eq:interac} gives the $m$-fold convolution of the
single-swimmer displacement \pdf.  This was the basis for the model used
in~\cite{Thiffeault2010b,Lin2011} and in an earlier version of this
paper~\cite{Thiffeault2014_preprint_v1}.  We have thus shown that
formula~\eqref{eq:rhotxNswim} is the natural infinite-volume limit of the
interaction picture.

Formula~\eqref{eq:interac} is very useful in many instances, such as
when~$\Mpalm$ is small, in which case only a few terms are needed
in~\eqref{eq:interac} for a very accurate representation.  Note that the first
term of the sum in~\eqref{eq:interac} is a $\delta$-function, which
corresponds to particles that are outside the interaction volume~$\Volint$.
This singular behavior disappears after $\prob_{\X_\pal}(\xc)$ is convolved
with a Gaussian distribution associated with molecular noise.

Let us apply~\eqref{eq:interac} to a specific example.  A model for cylinders
and spheres of radius~$\lsc$ traveling along the~$\zc$ axis in an inviscid
fluid~\cite{Thiffeault2010b,Lin2011} is the \emph{log model},
\begin{equation}
  \Delta_\pal(\etav) = 
  \begin{cases}
    \Clog\ln^+(\lsc/\rho),&\text{if $0 \le \zc \le \pal$,}\\
    0,&\text{otherwise},
  \end{cases}
  \label{eq:Deltalog}
\end{equation}
where~$\rho$ is the perpendicular distance to the swimming direction
and~$\ln^+\xc\ldef\ln\max(\xc,1)$.  The logarithmic form comes from the
stagnation points on the surface of the swimmer, which dominate transport in
this inviscid limit.  This model is also appropriate for a spherical
`treadmiller' swimmer in viscous flow.  The drift function~\eqref{eq:Deltalog}
resembles Fig.~\ref{fig:sphere_Delta}.

For the form~\eqref{eq:Deltalog} the interaction volume~$\Volint$ is the same
as~$\vpal$, the volume carved out during the swimmer's motion
(Eq.~\eqref{eq:vpal}).  By changing integration variable from~$\rho$
to~$\Delta$ in~\eqref{eq:cKdef} we can carry out the integrals explicitly to
obtain (see Appendix~\ref{apx:logmodel})
\begin{equation}
  \cKm_\pal(\kc) = \begin{cases}
    (1 + (\Clog\kc)^2)^{-1/2},
    \qquad &\text{(cylinders)}; \\
    (\Clog\kc/2)^{-1}\arctan(\Clog\kc/2),
    \qquad &\text{(spheres)}.
  \end{cases}
  \label{eq:cKlogmodel}
\end{equation}
This is independent of~$\pal$, even for short paths (but note
that~\eqref{eq:Deltalog} is not a good model for~$\pal < \lsc$).

Furthermore, for~$\sdim=2$ we can also explicitly obtain the convolutions that
arise in~\eqref{eq:interac} to find the full distribution,
\begin{equation}
  \prob_{\X_\pal}(\xc) = \ee^{-\Mpal}\l(\delta(\xc) + \sum_{\nenc=1}^\infty
  \frac{\Mpal^\nenc}{\nenc!}\,
    \frac{1}{\Clog\sqrt{\pi}\,\Gamma(\nenc/2)}
    \l(\lvert \xc\rvert/2\Clog\r)^{(\nenc-1)/2}
    K_{(\nenc-1)/2}(\lvert \xc\rvert/\Clog)\r),
    \label{eq:probNcyl}
\end{equation}
where~$K_\alpha(x)$ are modified Bessel functions of the second kind,
and~$\Gamma(x)$ is the Gamma function (not to be confused with~$\cK_\pal(\kc)$
above).  Equation~\eqref{eq:probNcyl} is a very good approximation to the
distribution of displacements due to inviscid cylinders.  Unfortunately no
exact form is known for spheres: we must numerically
evaluate~\eqref{eq:rhotxNswim} with~\eqref{eq:cKlogmodel} or use asymptotic
methods (see Section~\ref{sec:largedev}).

\section{Long paths: Large-deviation theory}
\label{sec:largedev}

In Section~\ref{sec:interact} we derived an alternative form of our master
equation~\eqref{eq:rhotxNswim} as an expansion in an `interaction' volume.
Here we look at another way to evaluate the inverse Fourier transform
in~\eqref{eq:rhotxNswim}, using large-deviation
theory~\cite{Gartner1977,Ellis1984,Ellis,Touchette2009}.  In
essence, large-deviation theory is valid in the limit when a particle
encounters many swimmers, so that~$\Mpal$ is large (in practice `large' often
means order one for a reasonable approximation).  This includes the central
limit theorem (Gaussian form) as a special case.  In this section we provide a
criterion for how much time is needed before Gaussian behavior is observed,
which can help guide future experiments.

Earlier we used the characteristic function~\eqref{eq:largevol}.  Here it is
more convenient to work with the moment-generating function, which in our case
can be obtained simply by letting~$\sc=\imi\kc$.  The moment-generating
function of the distribution is then
\begin{equation*}
  \avg{\ee^{\sc\X_\pal}} =
  \exp\l(-\Mpal\,\cK_\pal(-\imi\sc)\r) =
  \exp\l(\Mpal\,\Lpal(\sc)\r)
\end{equation*}
where~$\Mpal$ was defined by Eq.~\eqref{eq:Mpal}, and
\begin{equation}
  \Lpal(\sc) \ldef
  \frac{1}{\Mpal}\ln\avg{\ee^{\sc\X_\pal}} = -\cK_\pal(-\imi\sc)
\end{equation}
is the scaled cumulant-generating function.  As its name implies, this
function has the property that its derivatives at~$\sc=0$ give the cumulants
of~$\X_\pal$ scaled by~$\Mpal$, for example
\begin{equation}
  \Lpal''(0) = \Mpal^{-1}\avg{\X_\pal^2},
  \qquad
  \Lpal''''(0) = \Mpal^{-1}\l(\avg{\X_\pal^4} - 3\avg{\X_\pal^2}^2\r),
  \label{eq:cumul}
\end{equation}
where we left out the vanishing odd moments.  We left out the~$\pal$ subscript
on~$\Lpal(\sc)$ since we assume that it becomes independent of~$\pal$ for
large~$\pal$.

If~$\Lpal(\sc)$ is differentiable over some interval of
interest,~$\prob_{\X_\pal}(\xc)$ satisfies a \emph{large-deviation
  principle}~\cite{Gartner1977,Ellis1984,Ellis,Touchette2009},
\begin{equation}
  \prob_{\X_\pal}(\xc)
  \sim \ee^{-\Mpal\,\I(\xc/\Mpal) + \order{\Mpal}},\qquad
  \Mpal \gg 1,
  \label{eq:rholargedev}
\end{equation}
where~$\I(\xa)$ is the \emph{rate function}, which is the Legendre--Fenchel
transformation of~$\Lpal(\sc)$:
\begin{equation}
  \I(\xa) = \sup_{\sc\in \mathbb{R}}\{\sc\xa - \Lpal(\sc)\}.
  \label{eq:Idef}
\end{equation}
The large-deviation principle is in essence an application of the method of
steepest descent for large~$\Mpal$.

\begin{figure}
  \begin{center}
    \includegraphics[width=.6\textwidth]{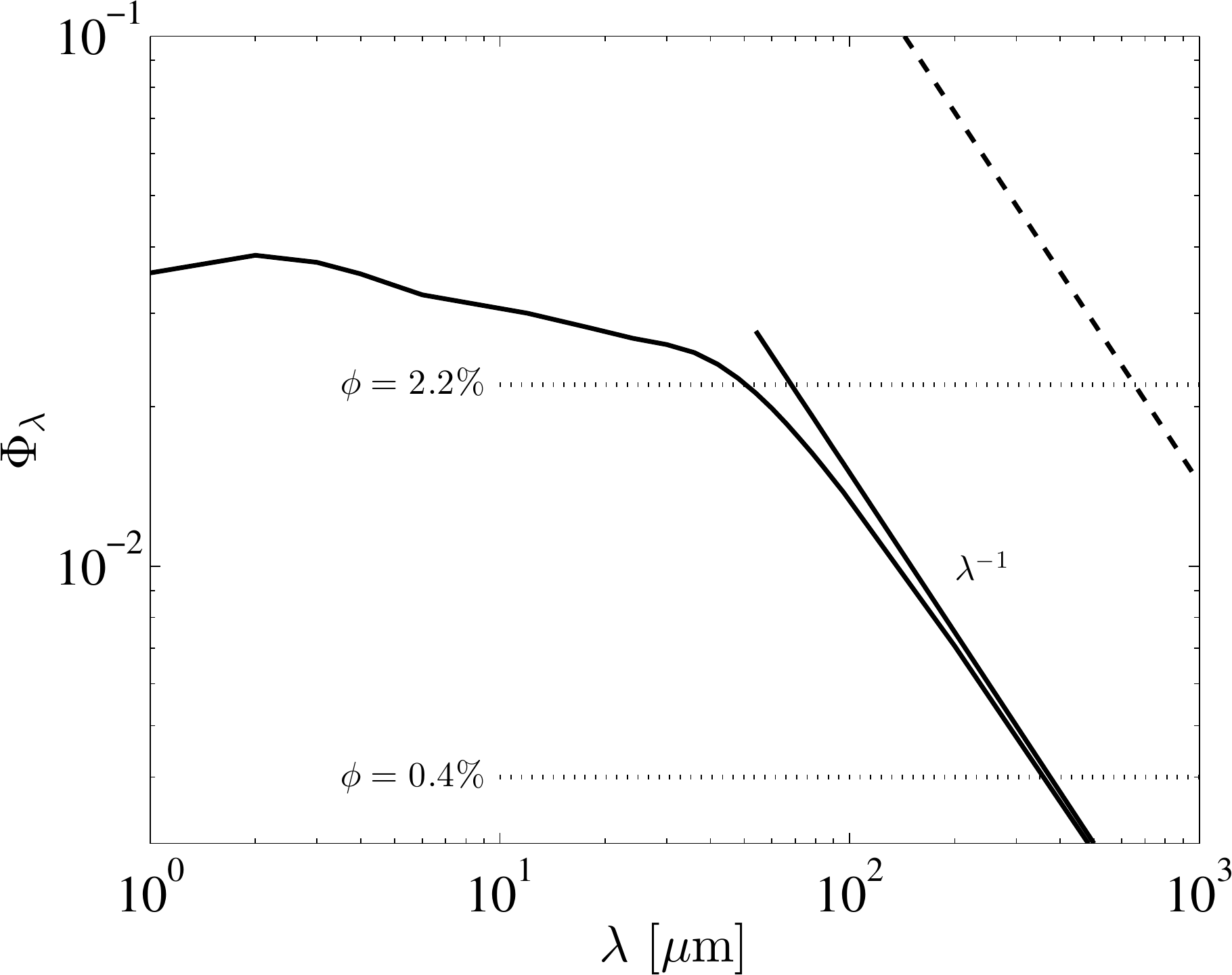}
  \end{center}
  \caption{The minimum volume fraction~$\Phi_\pal$ for the threshold of
    Gaussian behavior (Eq.~\eqref{eq:Gausscrit2}).  The solid line is the the
    squirmer model (Section~\ref{sec:numerics}) with~$\beta=0.5$ and
    radius~$\lsc=5\,\microm$.  The dashed line is for spherical treadmillers
    (inviscid spheres) of the same radius.  The latter require an order of
    magnitude longer to achieve Gaussianity, due to the short range of their
    velocity field.}
  \label{fig:Philambda}
\end{figure}

The scaled cumulant-generating function~$\Lpal(\sc)$ is always convex,
which guarantees a unique solution to~\eqref{eq:Idef}.  The rate
function~$\I(\xa)$ is also convex, with a global minimum at~$\xa=0$.  This
means that for small~$\xa=\xc/\Mpal$ we can use the Taylor expansion
\begin{equation}
  \I(\xa) = \tfrac12 \I''(0)\xa^2 + \tfrac1{4!} \I''''(0)\xa^4
  + \Order{\xa^6}
\end{equation}
to write
\begin{equation}
  \prob_{\X_\pal}(\xc)
  \sim \ee^{-\tfrac12\I''(0)\,\xc^2/\Mpal},\qquad
  \xc \ll \cc\,\Mpal,\quad
  \Mpal \gg 1,
  \label{eq:rhoGaussian}
\end{equation}
with~$\cc = \lvert12\I''(0)/\I''''(0)\rvert^{1/2}$.  This is a Gaussian
approximation with variance~$\Mpal/\I''(0)$, which can be shown to agree
with~\eqref{eq:r2N} after multiplying by~$\sdim$.  To recover a Gaussian
distribution over an appreciable range of~$\xc$ (say, a standard deviation) we
insert~$\xc \sim \sqrt{\Mpal/\I''(0)}$ in the condition~$\xc \ll \cc\,\Mpal$
to find the Gaussian criterion
\begin{equation}
  \Mpal \gg \frac{1}{12}\,\frac{\lvert\I''''(0)\rvert}{(\I''(0))^2}
  = \frac{1}{12}\,\frac{\lvert\Lpal''''(0)\rvert}{(\Lpal''(0))^2}.
  \label{eq:Gausscrit}
\end{equation}
After using~$\Lpal(\sc)$ to find the cumulants, we can rewrite this as
\begin{equation}
  \Phi_\pal \ldef
  \frac{(\sdim+3)}{40}\,
  \frac{\volsw\int_\Vol\Delta_\pal^4(\etav)\dint\Vol_{\etav}}
  {\l(\int_\Vol\Delta_\pal^2(\etav)\dint\Vol_{\etav}\r)^2}
  \ll \phi,
  \label{eq:Gausscrit2}
\end{equation}
where~$\volsw$ is the volume of one swimmer.  When~\eqref{eq:Gausscrit}
or~\eqref{eq:Gausscrit2} is satisfied, we can expect that the distribution
will be Gaussian (except in the far tails).  (The constant prefactor
in~\eqref{eq:Gausscrit2} is only valid for~$\sdim=2$ or~$3$.)  The
criterion~\eqref{eq:Gausscrit2} can be interpreted as the minimum volume
fraction~$\Phi_\pal$ required to observe Gaussian behavior, roughly within a
standard deviation of the mean.  We note that, at small swimmer volume
fraction, a long time (\ie, path length~$\pal$) is required to achieve the
Gaussian form.  Figure~\ref{fig:Philambda} highlights this: the solid curve
is~$\Phi_\pal$ from Eq.~\eqref{eq:Gausscrit2} for the squirmer model in
Section~\ref{sec:numerics}, with parameter values appropriate for the
experiments of \citet{Leptos2009}.  Their experiments had $\pal \lesssim
30\,\microm$, so they are in the slowly-decreasing region of
Fig.~\ref{fig:Philambda}, before more rapid~$\pal^{-1}$ convergence sets in
after~$\pal \gtrsim 50\,\microm$.  It is thus not surprising that Gaussian
tails were not observed in the experiments.

As an illustration of the large-deviation approach, we consider again the
inviscid cylinder and sphere results~\eqref{eq:cKlogmodel}.  We have then
respectively
\begin{equation}
  \Lpal(\sc) = \begin{cases}
    (1 - (\Clog\sc)^2)^{-1/2} - 1,
    \qquad &\text{(cylinders)}; \\
    (\Clog\sc/2)^{-1}\arctanh(\Clog\sc/2) - 1,
    \qquad &\text{(spheres)}.
  \end{cases}
  \label{eq:L2cylsph}
\end{equation}
We can see from~\eqref{eq:Idef} that the singularities in~\eqref{eq:L2cylsph}
($\lvert\sc\rvert = 1/\Clog$ for cylinders, $\lvert\sc\rvert = 2/\Clog$ for
spheres) immediately lead to~$\I(\xa) \sim \lvert\xa\rvert/\Clog$
and~$2\lvert\xa\rvert/\Clog$ as~$\lvert\xa\rvert \rightarrow \infty$,
respectively, corresponding to exponential tails in~\eqref{eq:rholargedev}
independent of~$\Mpal$.  These are the displacements of particles that come
near the stagnation points at the surface of the cylinder or
sphere~\cite{Lin2011}.  We can also use~\eqref{eq:L2cylsph} to compute the
constant on the right-hand side of~\eqref{eq:Gausscrit}: $3/4$ (cylinders)
and~$9/10$ (spheres), which are both of order unity.  This reflects the fact
that the drift function~$\Delta_\pal(\etav)$ is very localized, so convergence
to Gaussian is tied directly to the volume carved out by the swimmers.  For
swimmers with a longer-range velocity field, such as squirmers, the constant
is much larger, as reflected by the large difference between the solid
(squirmers) and the dashed (inviscid spheres) curves in
Fig.~\ref{fig:Philambda}.

\begin{figure}
  \begin{center}
    \includegraphics[width=.6\textwidth]{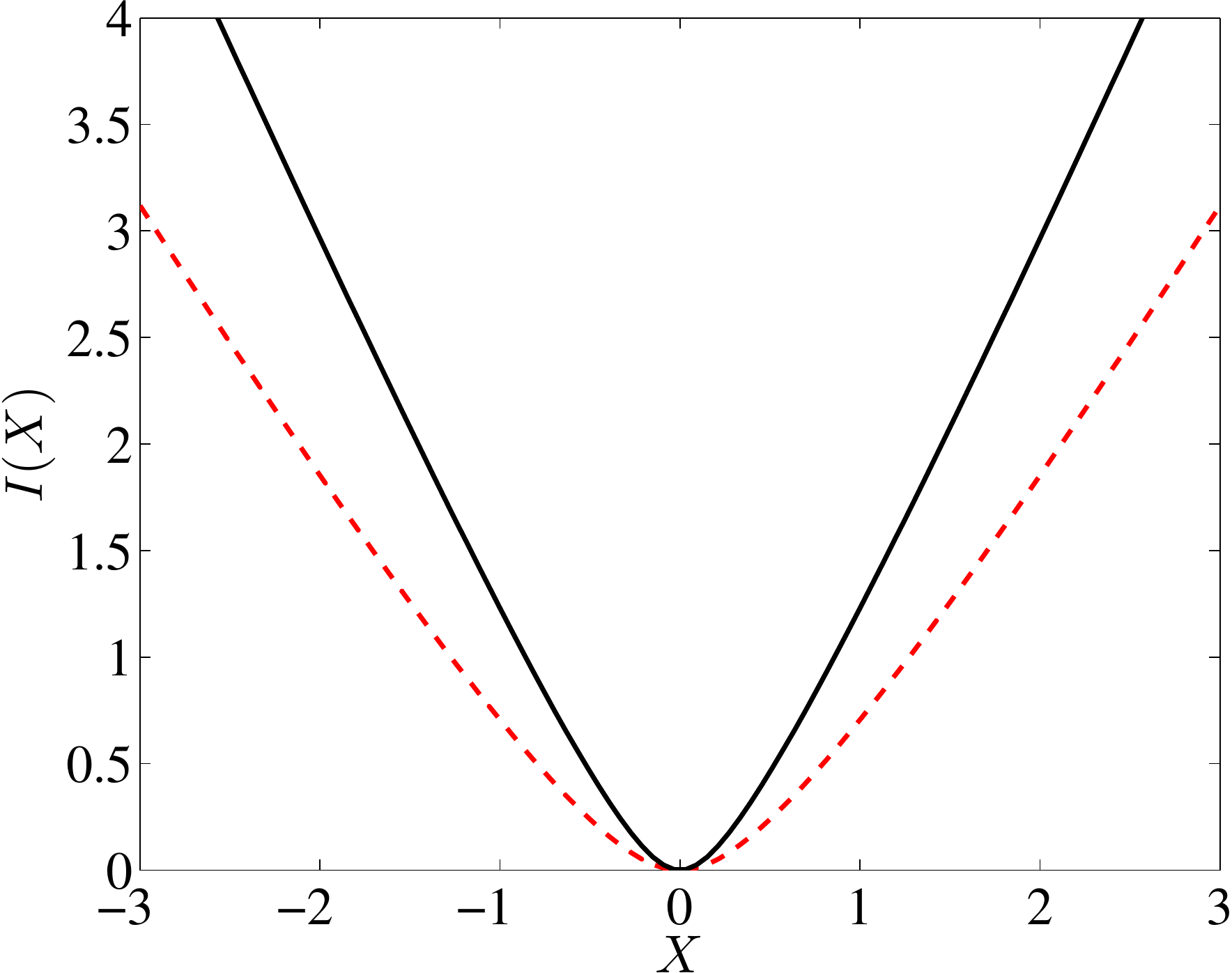}
  \end{center}
  \caption{The rate function~$\I(\xa)$ for cylinders (Eq.~\eqref{eq:Icyl},
    dashed line) and spheres (solid line, numerical solution
    of~\eqref{eq:rholargedev}).  In both cases we used~$\Clog=1$.  The linear
    behavior for large~$\lvert\xa\rvert$ indicates exponential tails
    in~\eqref{eq:rholargedev}.  When~$\xa$ is small, expanding near the
    quadratic minimum recovers the Gaussian limit.}
  \label{fig:cyl_sph_Cramer}
\end{figure}

For inviscid cylinders the Legendre--Fenchel transform~\eqref{eq:Idef} can be
done explicitly to find (with \hbox{$\Clog=1$})
\begin{equation}
  \I(\xa) =
  1 - \sqrt{3\pi\alpha}\l(12 - \alpha^2\xa^{-2}\r)^{-1/2}
  + \tfrac12\sqrt{\pi\alpha}
  \l(\l(\pi\alpha - 4\r)\alpha^{-2}\xa^2 + \tfrac13\r)^{1/2}
  \label{eq:Icyl}
\end{equation}
where~$\alpha(\xa) \ge 0$ is defined by
\begin{equation}
  \alpha^3(\xa) = 6\l(\sqrt{(9\pi\xa^4)^2 + 48\xa^6} - 9\pi\xa^4\r).
\end{equation}

For spheres~\eqref{eq:Idef} must be solved numerically for each~$\xa$, which
is straightforward since this is a one-dimensional problem with a unique
solution.  The function~$\I(\xa)$ for both cylinders and spheres is plotted in
Fig.~\ref{fig:cyl_sph_Cramer}.

\section{The diffusive scaling}
\label{sec:diffscal}

\begin{figure}
\begin{center}
\subfigure[]{
  \raisebox{.6em}
  {\includegraphics[height=.25\textheight]{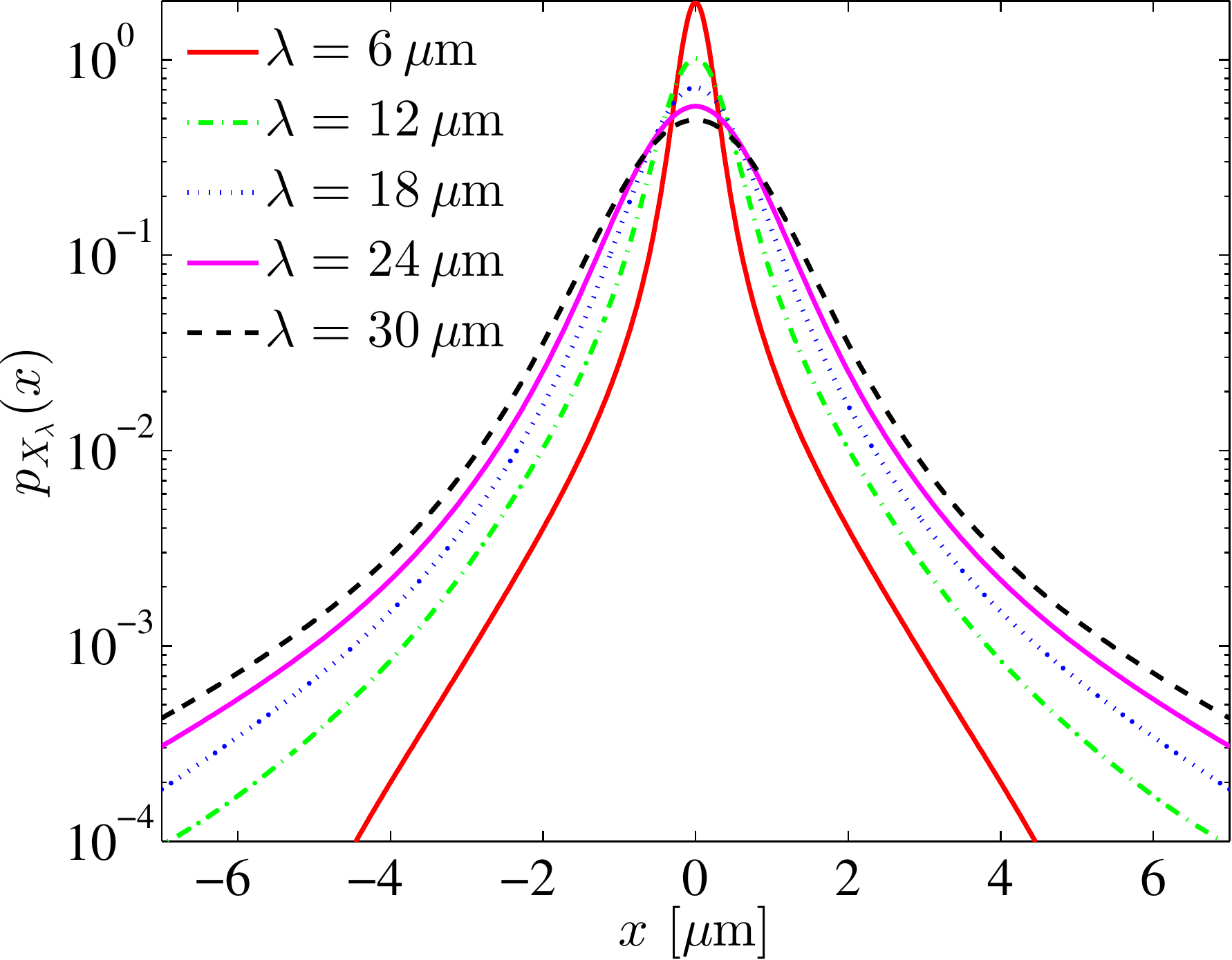}}
  \label{fig:diffusive_scaling_nodiff}
}
\hspace{1em}
\subfigure[]{
  \includegraphics[height=.26\textheight]{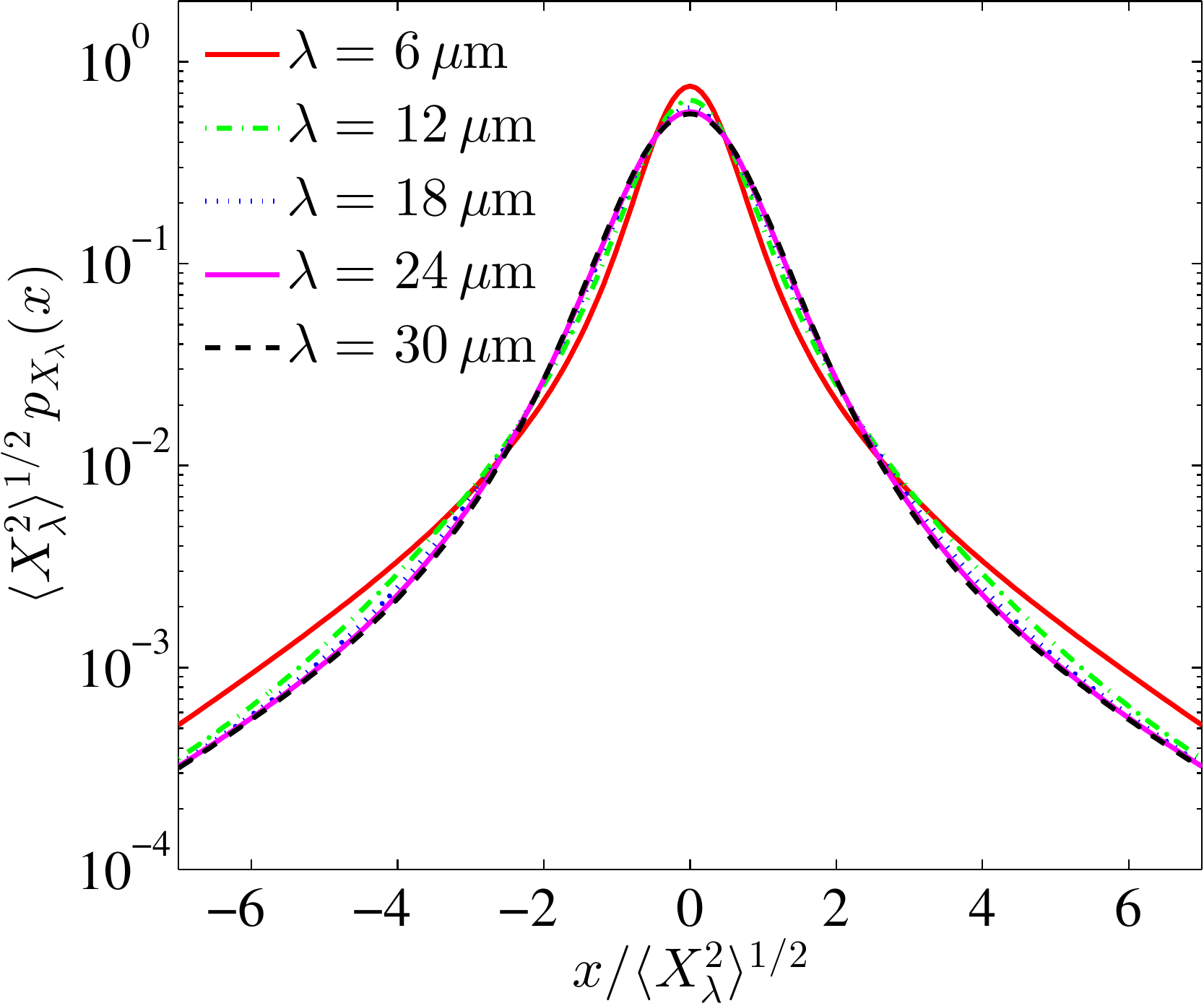}
  \label{fig:diffusive_scaling_nodiff_std1}
}
\end{center}
\caption{\protect\coloronline (a) \pdf{}s of particle displacements for
  squirmers for different times, at a number density~$\phi=2.2\%$.  (b) The
  same \pdf{}s rescaled by their standard deviation exhibit the `diffusive
  scaling' observed in the experiments of \citet{Leptos2009}, where the curves
  collapse onto one despite not being Gaussian.  As in the experiments, the
  scaling is worst for~$\pal=6\,\microm$.}
\label{fig:diffusive_scaling_nodiff_both}
\end{figure}

One of the most remarkable property of the \pdf{}s found by
\citeauthor{Leptos2009} is the \emph{diffusive scaling}.  This is illustrated
in Fig.~\ref{fig:diffusive_scaling_nodiff_both}: the unrescaled displacement
\pdf{}s are shown in Fig.~\ref{fig:diffusive_scaling_nodiff}; the same \pdf{}s
are shown again in Fig.~\ref{fig:diffusive_scaling_nodiff_std1}, but rescaled
by their standard deviation.  The \pdf{}s collapse onto a single curve (the
shortest path length collapses more poorly).
Figure~\ref{fig:diffusive_scaling_nodiff_both} was obtained in the same manner
as Fig.~\ref{fig:compare_to_Leptos_both}, using our probabilistic approach.
Hence, the diffusive scaling is also present in our model, as it was in the
direct simulations of \citet{Lin2011} for a similar range of path lengths.  In
Fig.~\ref{fig:diffusive_scaling_nodiff_both} we left out thermal diffusion
completely, which shows that it is not needed for the diffusive scaling to
emerge.

\begin{figure}
\begin{center}
\subfigure[]{
  \raisebox{.6em}
  {\includegraphics[height=.25\textheight]{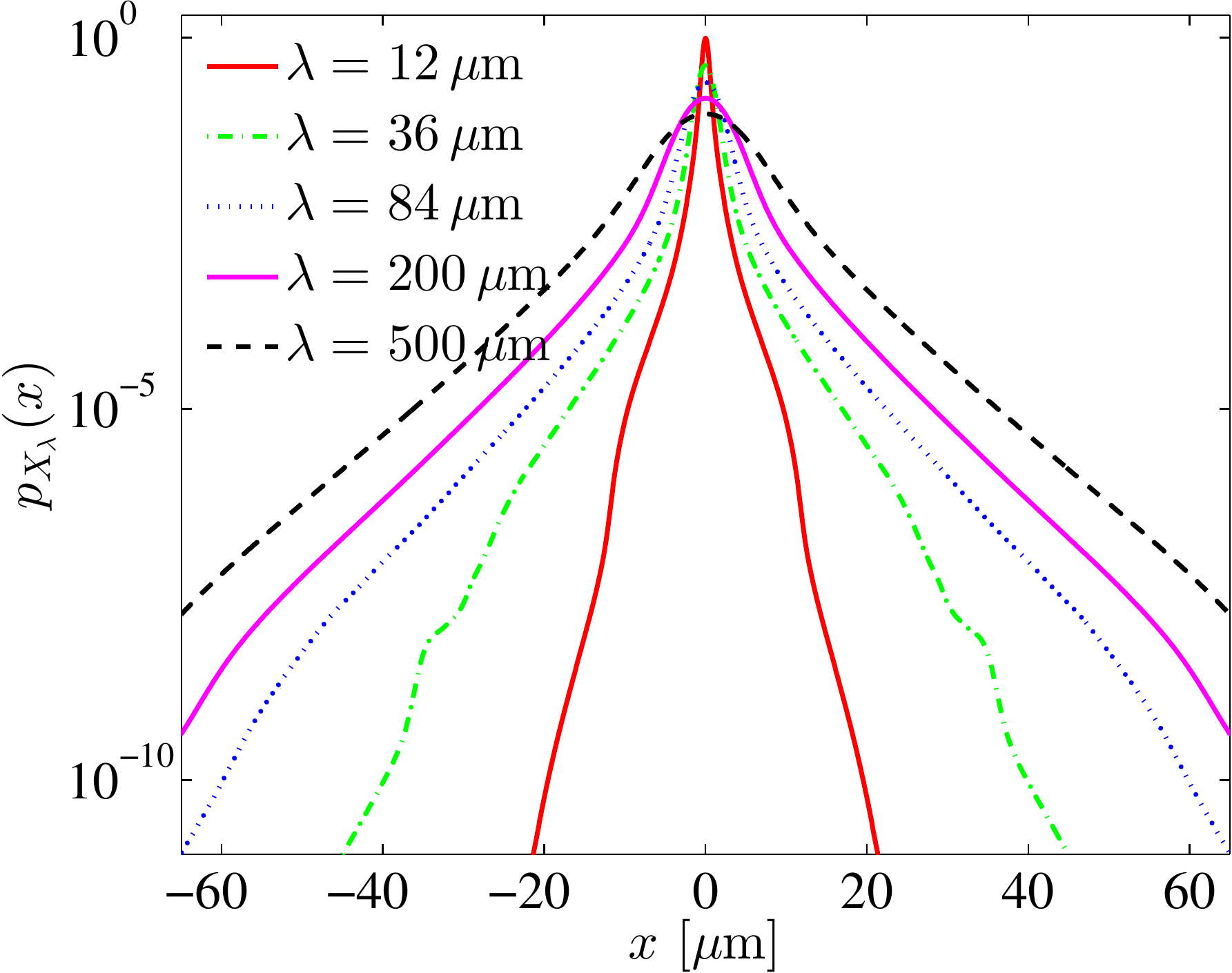}}
  \label{fig:diffusive_scaling_nodiff_full}
}
\hspace{1em}
\subfigure[]{
  \includegraphics[height=.26\textheight]{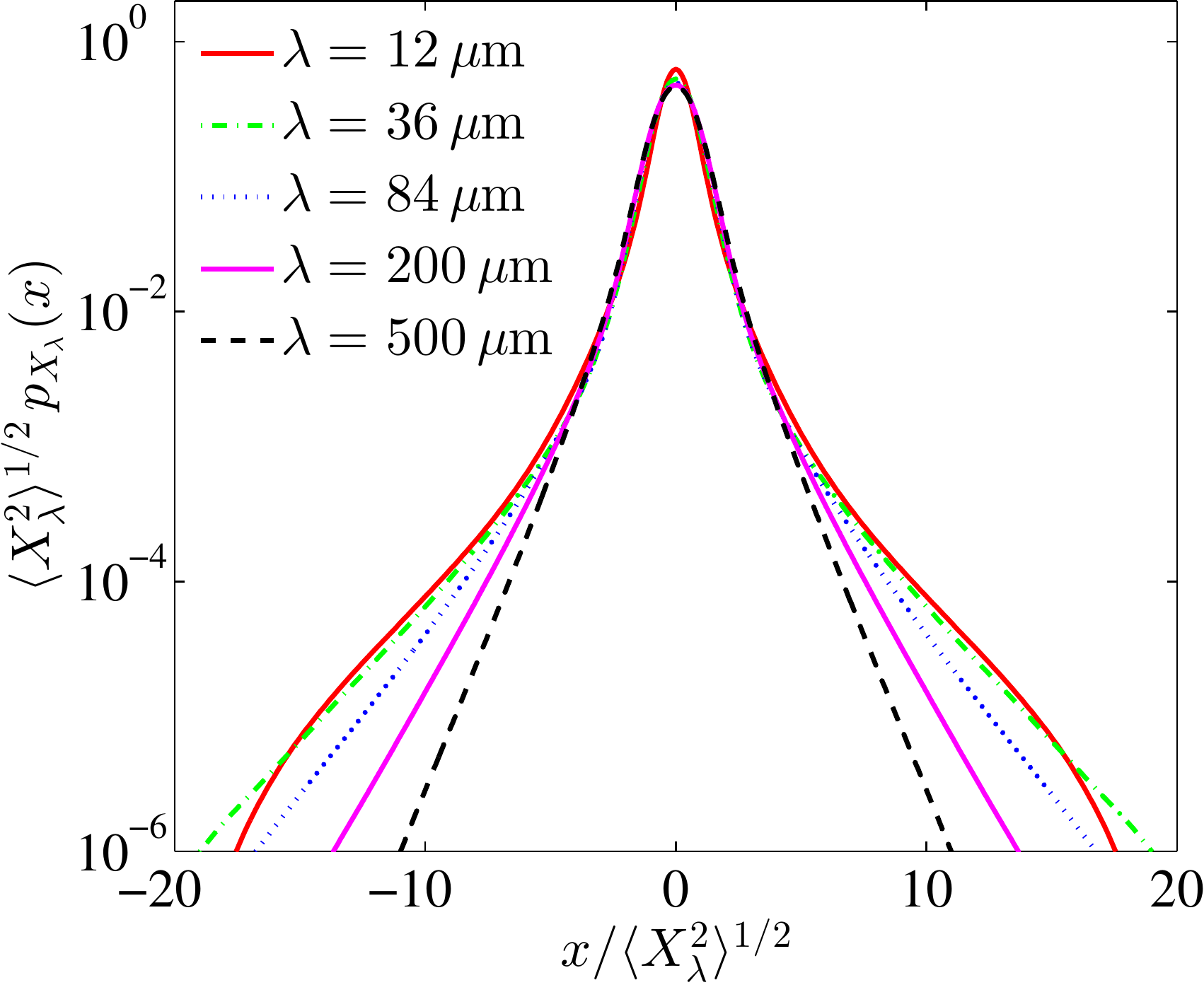}
  \label{fig:diffusive_scaling_nodiff_std1_full}
}
\end{center}
\caption{\protect\coloronline Same as
  Fig.~\ref{fig:diffusive_scaling_nodiff_both} but for longer times and with a
  wider scale.  In (a) the distributions broaden with time since their
  standard deviation is increasing; in (b), after rescaling by the standard
  deviation, the distributions' tails narrow with increasing~$\pal$ as they
  converge to a Gaussian.}
\label{fig:diffusive_scaling_nodiff_full_both}
\end{figure}

Here we have the luxury of going much further in time and to examine the
probability of larger displacements, since we are simply carrying out
integrals and not running a statistically-limited experiment or simulation.
(The numerical integrals are of course limited by resolution.)
Figure~\ref{fig:diffusive_scaling_nodiff_full_both} shows much longer runs
(maximum~$\pal=500\,\microm$ compared to~$30\,\microm$ in the experiments).
We see that, though the diffusive scaling holds in the core (as it must, since
the core is Gaussian), the tails are narrowing, consistent with convergence to
a Gaussian distribution but breaking the diffusive scaling.  We now explain
why the diffusive scaling appears to hold for some time, but eventually breaks
down.

\begin{figure}
  \begin{center}
    \includegraphics[width=.45\textwidth]{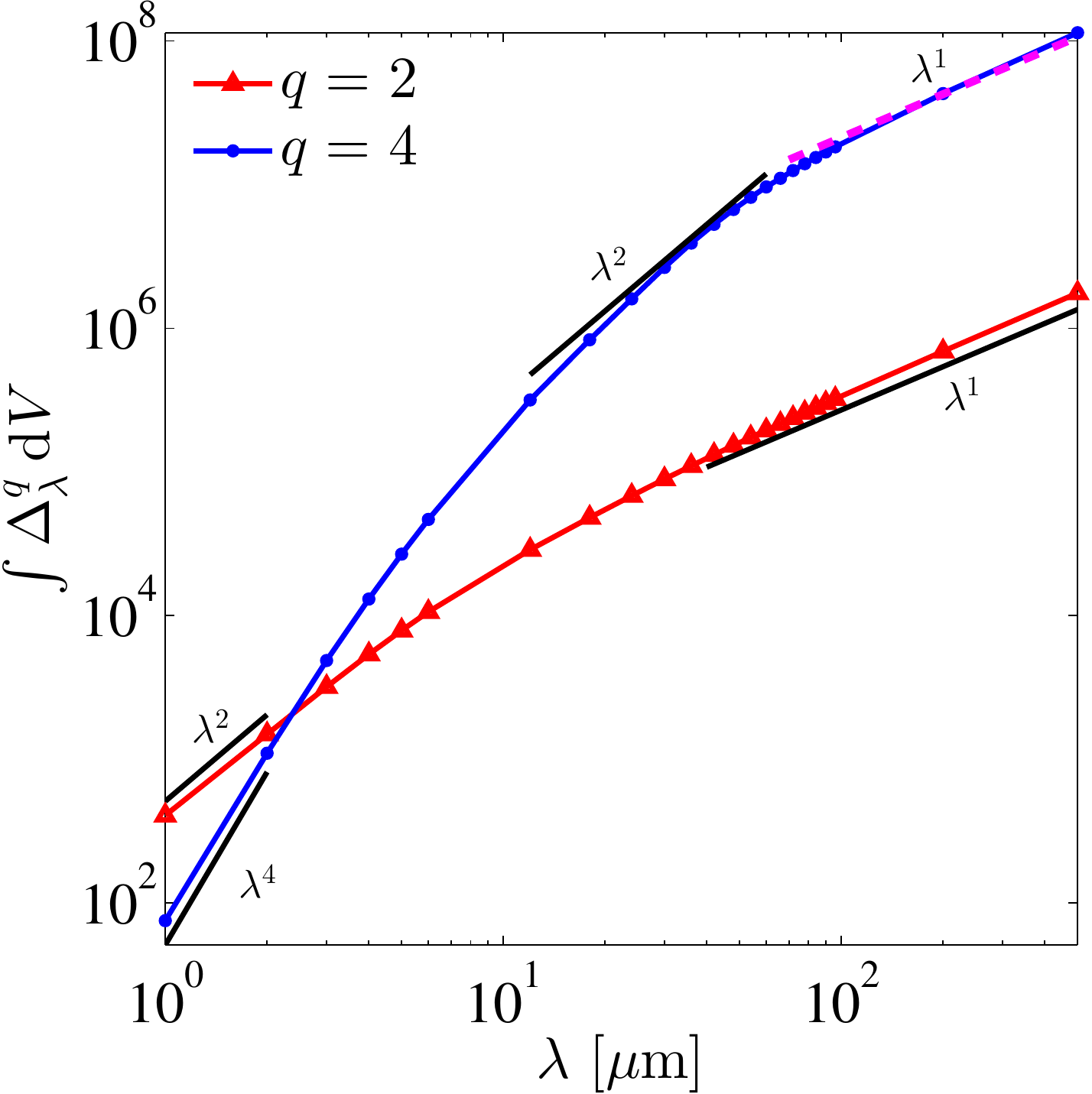}
  \end{center}
  \caption{\protect\coloronline The second and fourth integrated moments
    of~$\Delta_\pal$.  These grow ballistically ($\pal^q$) for short times,
    and eventually grow linearly with~$\pal$.  The slow crossover
    of~$\Delta_\pal^4$ is the origin of the `diffusive scaling' of
    \citet{Leptos2009}, since in their narrow range of~$\pal$ the curve is
    tangent to~$\pal^2$.}
  \label{fig:intDeltaq}
\end{figure}

To understand the origin of the diffusive scaling, let us first examine how
the integrated moments of~$\Delta_\pal$ change with~$\pal$.
Figure~\ref{fig:intDeltaq} shows the evolution of the spatial integrals
of~$\Delta_\pal^2$ and~$\Delta_\pal^4$ for our squirmer model.  For
short~$\pal$, the moment of~$\Delta_\pal^q$ grows as~$\pal^q$.  This is a
typical `ballistic' regime: it occurs because for short times the integrals
are dominated by fluid particles that are displaced proportionately to the
swimmer's path length.  These particles are typically very close to the
swimmer, and get dragged along for a while.  This regime is visible for~$\pal
\lesssim 2\,\microm$ in Fig.~\ref{fig:intDeltaq}.

As~$\pal$ becomes larger, the particles initially near the swimmer are left
behind, and thus undergo only a finite displacement even as~$\pal$ increases.
Eventually, for~$q=2$ the scenario illustrated in
Fig.~\ref{fig:stresslet_Delta} takes over and leads to linear growth of the
moment with~$\pal$.  This can be seen in Fig.~\ref{fig:intDeltaq} (triangles)
for~$\pal \gtrsim 40\,\microm$, though the scaling already looks fairly linear
at~$\pal \sim 10\,\microm$.  For~$q=4$ the moment also eventually grows
linearly with~$\pal$, but the mechanism is different: the larger power
downplays the far-field stresslet effect, and the near-field dominates.  The
linear growth is thus due to a corresponding linear growth of the support
of~$\Delta_\pal^4$ as in Fig.~\ref{fig:sphere_Delta}.  This can be seen in
Fig.~\ref{fig:intDeltaq} (dots) for~$\pal \gtrsim 100\,\microm$, as indicated
by a dashed line (see Appendix~\ref{apx:logmodel} for the computation of this
asymptotic form).  The crucial fact is that for~$q=4$ the crossover
from~$\pal^q$ to~$\pal^1$ takes much longer than for~$q=2$.  This is because
the larger power weighs the largest displacements (with~$\Delta_\pal^q \sim
\lambda^q$) more heavily, so they dominate for longer before becoming too
rare.  This crossover is at the heart of the diffusive scaling, as we now
show.

Let us assume that the distribution~$\prob_{\X_\tt}(\xc)$ does satisfy a
diffusive scaling, such that $\alsct\,\prob_{\X_\pal}(\alsct\tilde\xc) =
\tilde\prob_{\X_\pal}(\tilde\xc)$ is independent of~$\pal$.
From~\eqref{eq:rhotxNswim}, after changing integration variable to~$\tilde\kc
= \alsct\kc$,
\begin{equation}
  \tilde\prob_{\X_\pal}(\tilde\xc)
  =
  \alsct\,\prob_{\X_\pal}(\alsct\tilde\xc)
  =
  \frac{1}{2\pi}\int_{-\infty}^\infty
  \exp\l(-\Mpal\,\cK_\pal(\tilde\kc/\alsct)\r)
  \ee^{-\imi\tilde\kc\tilde\xc}\dint\tilde\kc.
\end{equation}
Hence, a diffusive scaling law requires that $\Mpal\cK_\pal(\tilde\kc/\alsct)$
be independent of~$\pal$.  Using this scaling in~\eqref{eq:cKdef}, we have
\begin{equation}
  \Mpal\cK_\pal(\tilde\kc/\alsct) =
  \nd
  \int_\Vol\K(\Delta_\pal(\etav)\tilde\kc/\alsct)\dint\Vol_{\etav}\,.
\end{equation}
We Taylor expand~$\K$ (for~$\sdim=3$):
\begin{equation}
  \Mpal\cK_\pal(\tilde\kc/\alsct)/\nd
  =
  \tfrac{1}{6}\,\tilde\kc^2\pal^{-1}
  \int_\Vol\Delta_\pal^2(\etav)\dint\Vol_{\etav}
  +
  \tfrac{1}{120}\,\tilde\kc^4\pal^{-2}
  \int_\Vol\Delta_\pal^4(\etav)\dint\Vol_{\etav}
  + \Order{\tilde\kc^6}.
\end{equation}
The first term recovers the Gaussian approximation; the second is the first
correction to Gaussian.  Again this must be independent of~$\pal$ to obtain a
diffusive scaling, so we need
\begin{equation}
  \int_\Vol\Delta_\pal^2(\etav)\dint\Vol_{\etav} \sim \pal,
  \qquad
  \int_\Vol\Delta_\pal^4(\etav)\dint\Vol_{\etav} \sim \pal^2,
\end{equation}
and clearly in general we would need each even moment~$q$ to scale
as~$\pal^{q/2}$.  However, we've already seen that all the moments typically
eventually scale linearly with~$\pal$, so there can be no diffusive scaling.
Because there is a transition from a power larger than~$2$ ($\pal^4$) to one
less that~$2$ ($\pal^1$), observe that in Fig.~\ref{fig:intDeltaq} there is a
range of~$\pal$ (roughly $10\,\microm \lesssim \pal \lesssim 60\,\microm$)
where~$\pal^2$ is tangent to the~$q=4$ curve, as indicated by the line
segment.  In that range the distribution will appear to have a reasonably good
diffusive scaling, consistent with
Fig.~\ref{fig:diffusive_scaling_nodiff_both}.  But, as we saw in
Fig.~\ref{fig:diffusive_scaling_nodiff_full_both}, the diffusive scaling does
not persist for larger~$\pal$.  It is a coincidence that the range of~$\pal$
used in the experiments of \citet{Leptos2009} were exactly in that
intermediate regime.

\section{Discussion}
\label{sec:discussion}

In this paper, we showed how to use the single-swimmer drift function to fully
derive the probability distribution of particle displacements.  We took the
limit of infinite volume and discussed the underlying assumptions, such as the
need for the function~$\cK_\pal(\kc)$ in~\eqref{eq:propcond2} to not diverge
too quickly with volume.  In typical cases, the function becomes independent
of volume as we make~$\Vol$ large, but it is possible for the integral to
diverge with~$\Vol$, as may occur for example in sedimentation problems.  If
the divergence is rapid enough a larger value of~$\Mprop$ would need to be
used when applying Proposition~\ref{prop:expid}, potentially leading to
interesting new distributions.  Whether this can happen in practice is a topic
for future investigation.

An intriguing question is: why does the squirmer model do so well?  As was
observed previously~\cite{Lin2011,Pushkin2014}, it reproduces the \pdf\ very
well in the core and part of the tails (Fig.~\ref{fig:compare_to_Leptos}).
However, the high precision of our calculation reveals that the experiments
have slightly `fatter' tails.  This means that the specific details of the
organisms only begin to matter when considering rather large displacements.
In future work, we shall attempt to determine what is the dominant cause of
large displacements in the near-field for a more realistic model of
\textit{C.~reinhardtii}.  The large displacements could arise, for instance,
from the strong time-dependence of the swimming organism, or from particles
`sticking' to the no-slip body of the organism or to stagnation points.

We have not discussed at all the role of reorientation, that is,
running-and-tumbling or orientation diffusion.  \citet{Pushkin2013b} showed
that some curvature in the paths does not influence the diffusivity very much,
so it is likely not a very important factor here.  In experiments involving
different organisms it could matter, especially if the swimmer carries a
volume of trapped fluid.

One glaring absence from the present theory is any asymmetry between pushers
and pullers.  This suggests that correlations between swimmers must be taken
into account to see this asymmetry emerge.  These correlations begin to matter
as swimmer densities are increased.  However, how to incorporate these
correlations into a model similar to the one presented here is a challenge.

\begin{acknowledgments}
  The author thanks Bruno Eckhardt and Stefan Zammert for helpful discussions
  and for providing the digitized data from \citeauthoreos{Leptos2009}.  The
  paper also benefited from discussions with Raymond Goldstein, Eric Lauga,
  Kyriacos Leptos, Peter Mueller, Tim Pedley, Saverio Spagnolie, and Benedek
  Valko.  Much of this work was completed while the author was a visiting
  fellow of Trinity College, Cambridge. This research was supported by NSF
  grant DMS-1109315.
\end{acknowledgments}

\bibliography{bib/journals_abbrev,bib/articles}

\appendix

\section{Proof of Proposition~\ref{prop:expid}}
\label{apx:proof}

\begin{proof} Observe that~$\eprop\yprop(\eprop) \sim
  \order{\eprop^{1/(\Mprop+1)}} \rightarrow 0$ as~$\eprop \rightarrow 0$.
  Writing~$(1 - \eprop\yprop)^{1/\eprop} =
  \ee^{\eprop^{-1}\,\ln(1-\eprop\yprop)}$, we expand the exponent as a
  convergent Taylor series:
\begin{align*}
  (1 - \eprop\yprop)^{1/\eprop}
  &= \exp\biggl(-\eprop^{-1}\sum_{m=1}^{\infty}\frac{(\eprop\yprop)^m}{m}\biggr)
  \quad
  \text{(converges since $\eprop\yprop \sim \order{\eprop^{1/(\Mprop+1)}}$)} \\
  &= \exp\biggl(-\eprop^{-1}\biggl(\sum_{m=1}^{\Mprop}\frac{(\eprop\yprop)^m}{m}
  + \Order{(\eprop\yprop)^{\Mprop+1}}\biggr)\biggr) \\
  &= \exp\biggl(-\eprop^{-1}\sum_{m=1}^{\Mprop}\frac{(\eprop\yprop)^m}{m}\biggr)
  \exp\l(\Order{\eprop^\Mprop\yprop^{\Mprop+1}}\r) \\
  &= \exp\biggl(-\eprop^{-1}\sum_{m=1}^{\Mprop}\frac{(\eprop\yprop)^m}{m}\biggr)
  \l(1 + \order{\eprop^0}\r).
  \qedhere
\end{align*}
\end{proof}

\section{The log model}
\label{apx:logmodel}

A reasonable model for objects in an inviscid
fluid~\cite{Thiffeault2010b,Lin2011} is the drift function
\begin{equation}
  \Delta_\pal(\rho,\zc) = 
  \begin{cases}
    \widetilde{\Delta}(\rho),&\text{if $0 \le \xc \le \pal$,}\\
    0,&\text{otherwise},
  \end{cases}
  \label{eq:Deltaprelog}
\end{equation}
where~$\rho$ is the perpendicular distance to the swimming direction.  The
integral~\eqref{eq:cKdef} then simplifies to
\begin{equation}
  \cK_\pal(\kc) =
  \frac{1}{\vpal}
  \int_\Vol\K(\kc\Delta_\pal(\etav))\dint\Vol_{\etav}
  =
  \frac{(\sdim-1)}{\lsc^{\sdim-1}}
  \int_0^\infty\K(\kc\widetilde{\Delta}(\rho))\,\rho^{\sdim-2}\dint\rho.
\end{equation}
Assume a monotonic relationship between~$\rho$ and~$\widetilde{\Delta}(\rho)$;
we can then change the integration variable to~$\widetilde{\Delta}$:
\begin{equation}
  \cK_\pal(\kc)
  = \frac{(\sdim-1)}{\lsc^{\sdim-1}}\,\int_{0}^\infty\K(\kc\widetilde{\Delta})
  \,\rho^{\sdim-2}(\widetilde{\Delta})
  \,\frac{\!\dint\widetilde{\Delta}}
  {\lvert\widetilde{\Delta}'(\rho(\widetilde{\Delta}))\rvert}\,.
  \label{eq:theint}
\end{equation}
To be more specific, let us use the \emph{log model}:
\begin{equation}
  \widetilde{\Delta}(\rho) = \Clog\ln^+(\lsc/\rho),
  \label{eq:logmodel}
\end{equation}
where~$\ln^+\xc\ldef\ln\max(\xc,1)$.  Here~$\widetilde{\Delta} \in [0,\infty)$
for~$\rho \in (0,\lsc]$.  The constant~$\Clog$ is set by the linear structure
of the stagnation points around the
swimmer~\cite{Thiffeault2010b,GFD2010_JLT_lect3,Lin2011}, and usually scales
with the size of the organism (\emph{not} the path length~$\pal$).  For
example,~$\Clog=\lsc$ for a cylinder of radius~$\lsc$ moving through inviscid
fluid~\cite{Thiffeault2010b,GFD2010_JLT_lect3}.  For spheres in the same type
of fluid,~$\Clog = \tfrac43\lsc$~\cite{Thiffeault2010b}.

We can write~$\rho = \lsc\,\ee^{-\widetilde\Delta/\Clog}$,
with~$\widetilde{\Delta}'(\rho) = \Clog/\rho =
(\Clog/\lsc)\,\ee^{\widetilde\Delta/\Clog}$.  The integral~\eqref{eq:theint}
is then
\begin{equation}
  \cK_\pal(\kc)
  = \frac{(\sdim-1)}{\Clog}
  \int_{0}^\infty\K(\kc\widetilde{\Delta})
  \,\ee^{-(\sdim-1)\widetilde\Delta/\Clog}
  \dint\widetilde{\Delta}.
\end{equation}
This is easily integrated to give~\eqref{eq:cKlogmodel}, after
using~\eqref{eq:cKm} with~$\Volint = \vpal$.

The log model is also appropriate for squirmers when computing
moments~$\int_\Vol\Delta_\pal^q\dint\Vol$ for~\hbox{$q > 2$}.  The
constant~$\Clog$ of Eq.~\eqref{eq:logmodel} is then~$\Clog(\beta) =
\tfrac43(1-\beta^2)^{-1}\lsc$, obtained by linearization around the two
stagnation points at the front and rear of the squirmer
(Fig.~\ref{fig:squirmer_contour_beta=0p5}).  For~$\beta^2 \ge 1$ the topology
of the stagnation points changes and this expression becomes invalid --- the
squirmer develops a trapped `bubble' or wake~\cite{Lin2011}.  To get a
reasonably accurate representation of~$\Delta_\pal$ it is important to also
include the constant correction to the log approximation.  This constant can
be absorbed in the choice of~$\lsc$ in~\eqref{eq:logmodel}, instead of using
the swimmer radius.  This gives an `effective radius'~$\lsc \rightarrow
\bb(\beta)\lsc$ in~\eqref{eq:logmodel}, which for our squirmer is smaller than
the `true' swimmer radius~$\lsc$.  The explicit calculation of this correction
involves thorny integrals and is beyond the scope of this paper.  The relevant
numerical values for our purposes are~$\bb(0.5)=0.5382$ and~$\bb(0)=0.6464$
(inviscid sphere limit~\cite{Thiffeault2010b}).  The log model was used to
compute the asymptotic form (dashed line) in Fig.~\ref{fig:intDeltaq}.

\end{document}